# The photochemical ring-opening reaction of 1,3-cyclohexadiene: complex dynamical evolution of the reactive state

O. Travnikova[1,#], T. Piteša[2,#], A. Ponzi[2], M. Sapunar[2], R. J. Squibb[3], R. Richter[4], P. Finetti[4], M. Di Fraia[4], A. De Fanis[5], N. Mahne[6], M. Manfredda[4], V. Zhaunerchyk[3], T. Marchenko[1], R. Guillemin[1], L. Journel[1], K. C. Prince[4], C. Callegari[4], M. Simon[1], R. Feifel[3], P. Decleva[7], N. Došlić[2,*] and M.N. Piancastelli[1,8,*]

[1]*Sorbonne Université, CNRS, Laboratoire de Chimie Physique-Matière et Rayonnement, LCPMR, F-75005, Paris, France*
[2]*Institut Ruđer Bošković, HR-10000 Zagreb, Croatia*
[3]*Department of Physics, University of Gothenburg, SE-412 96 Gothenburg, Sweden*
[4]*Elettra-Sincrotrone Trieste, 34149 Basovizza, Trieste, Italy*
[5]*European XFEL, D-22869 Schenefeld, Germany.*
[6]*IOM-CNR, S.S. 14 km 163.5 in Area Science Park, 34149 Trieste, Italy*
[7]*Dipartimento di Scienze Chimiche e Farmaceutiche, Universitá di Trieste, I-34127 Trieste, Italy.*
[8]*Department of Physics and Astronomy, Uppsala University, SE-751 20 Uppsala, Sweden*

Abstract

The photochemically induced ring-opening isomerization reaction of 1,3-cyclohexadiene (CHD) to 1,3,5-hexatriene (HT) is a textbook example of a pericyclic reaction, and has been amply investigated with advanced spectroscopic techniques. The generally accepted description of the isomerization pathway starts with a valence excitation to the lowest-lying bright state, followed by a passage through a conical intersection to a dark doubly excited state, and finally a branching between either the return to the ground state of the cyclic molecule or the actual ring-opening reaction leading to the open-chain isomer. It was traditionally assumed that the dark reactive state corresponds to the second excited state of CHD at the Franck-Condon geometry. Here in a joint experimental and computational effort we demonstrate that the evolution of the excitation-deexcitation process is much more complex than usually described. In particular, we show that an initially high-lying electronic state smoothly decreasing in energy along the reaction path plays a key role in the ring-opening reaction. The conceptual basis of our work is that the dynamics to consider here is determined by diabatic states, whose populations are the ones closely related to the observed photoelectron signal.

[#] These authors have contributed equally

*To whom correspondence should be addressed, maria-novella.piancastelli@physics.uu.se, nadja.doslic@irb.hr

The photochemical ring-opening reaction of 1,3-cyclohexadiene (CHD) to 1,3,5-hexatriene (HT) is a textbook example of a pericyclic reaction [1,2]. In addition to many theoretical investigations [3–12], the dynamical evolution of photoexcited CHD has been studied with a large variety of spectroscopic techniques, possibly making this photochemical reaction the most explored with isolated-molecule fundamental methods, i.e. in collision-free conditions and without solvent effects (see e.g. [13–19]). This pronounced interest arises from the fundamental importance of the reaction, its biological relevance [20,21] and a range of applications in organic synthesis and materials science [22–26].

In general, photochemical reactions are known to be driven by conical intersections (CoIns). These are geometry points at which the energy separation between two adiabatic Born–Oppenheimer potential energy surfaces (PESs) becomes smaller or comparable to the non-adiabatic couplings between these states, and which act as effective funnels for transfer of population between different adiabatic PESs [27-29]. Near a CoIn, the adiabatic PESs, which are the eigenvalues of the electronic Hamiltonian and thus the result of electronic structure calculations, conserve their energetic ordering but not their chemical character. PESs that retain their character and cross at CoIns are known as the diabatic states. Since properties such as the electronic transition dipole moment change smoothly only in the diabatic representation, the dynamics of the diabatic electronic populations is the one monitored in time-resolved ultrafast spectroscopy [30].

The conceptual framework to understand the photochemistry of CHD is provided by the Woodward–Hoffmann rules [31], extended by van der Lugt and Oosterhoff [32,33] and stating that the conrotatory ring opening reaction is mediated by a doubly excited electronic state of the same symmetry as the ground state ($1^1A^-$). In the ring-opening reaction the doubly occupied LUMO orbital of the reactive state becomes the doubly occupied HOMO of HT. The generally accepted sequence of events starts with a valence excitation by a wavelength of about 267 nm to the first ($S_1$) bright state, labelled $1^1B$, followed by a passage through a CoIn to a dark state, labelled $2^1A^-$. Notice that the labels indicate diabatic states and that we use the notation for alternant π-systems (*plus* and *minus*) to distinguish allowed (forbidden) transitions from the ground state to states of different (same) pseudosymmetry [34]. The following step is a branching between two pathways at a second CoIn, either the return to the ground state of the cyclic molecule or the actual ring-opening reaction leading to the open-chain isomer (see e.g. [11-14]).

Despite tremendous efforts, a conclusive proof of the reactivity of the diabatic $2^1A^-$ state has not yet been achieved. To address this problem, we ask a seemingly simple question: is there any other electronic state of symmetry A and partial double excitation character that may be involved in the ring-opening reaction in CHD? Here in a joint experimental and computational effort we show that the evolution of the process is more complex than usually described. In particular, we show that a high-lying state with a pronounced double-excitation character, labeled $2^1A^+$, plays a key role in the ring-opening reaction.

To observe the first stages of a photochemical process, a suitable method is to prepare a photoexcited state with an optical laser, the so-called pump, and then follow its evolution by valence photoelectron spectroscopy, the so-called probe, as a function of pump-probe time delay with resolution on the picosecond (ps, $10^{-12}$ s) or femtosecond (fs, $10^{-15}$ s) time scale. Time-resolved photoemission is the first-choice technique to follow the evolution of a system since it provides information on both electronic and nuclear dynamics. Furthermore, it provides information on states which are not reachable by absorption methods, in particular dark states which need to be characterized in the present case.

A breakthrough in this direction is represented by the FERMI free-electron laser (FEL) at the Elettra facility, Trieste, Italy. Time-resolved photoemission spectra can be obtained there with spectral resolution high enough to precisely characterize ionization from electronic states even if they are weak and/or close in energy (see e.g. [35,36]). Our experiments were performed there on the beam line LDM, devoted to atomic and molecular spectroscopy studies [37]. The pump was a titanium-sapphire optical laser, providing a wavelength of 267 nm, and the probe was valence photoelectron spectroscopy with a photon energy of 19.23 eV. The delay time range was from -1 to 2 ps, spanned in steps of 50-100 fs. We recorded valence photoelectron spectra with a magnetic bottle spectrometer [38,39] (see Supplementary Information, SI, for further details on the facility, the beam line and the spectrometer).

In Figure 1, upper panels (experiment and theory) we show valence photoelectron spectra recorded for several values of pump-probe delay in the delay range from -1 to 2 ps in steps of 100 fs. The spectra are plotted as a two-dimensional map, highlighting the variations in spectral intensity as a function of time delay. The ground-state spectrum is subtracted from the spectra obtained at later times (see SI and Supplementary Figs.2-4 for further details on how the experimental spectra are obtained). The upper panels illustrate the variation of the photoionization signal as a function of the pump-probe delay, the changes of averaged signals in the four characteristic areas are shown in the lower panels (experiment and theory). We notice the development of a series of new features in two electron energy regions, i.e. at low binding energy (4-8 eV) and around 10 eV. The low-binding-energy region is characteristic of spectral features related to excited states or dynamic features due to the photoexcitation process, while the 10 eV region is in the range characteristic of ground-state features.

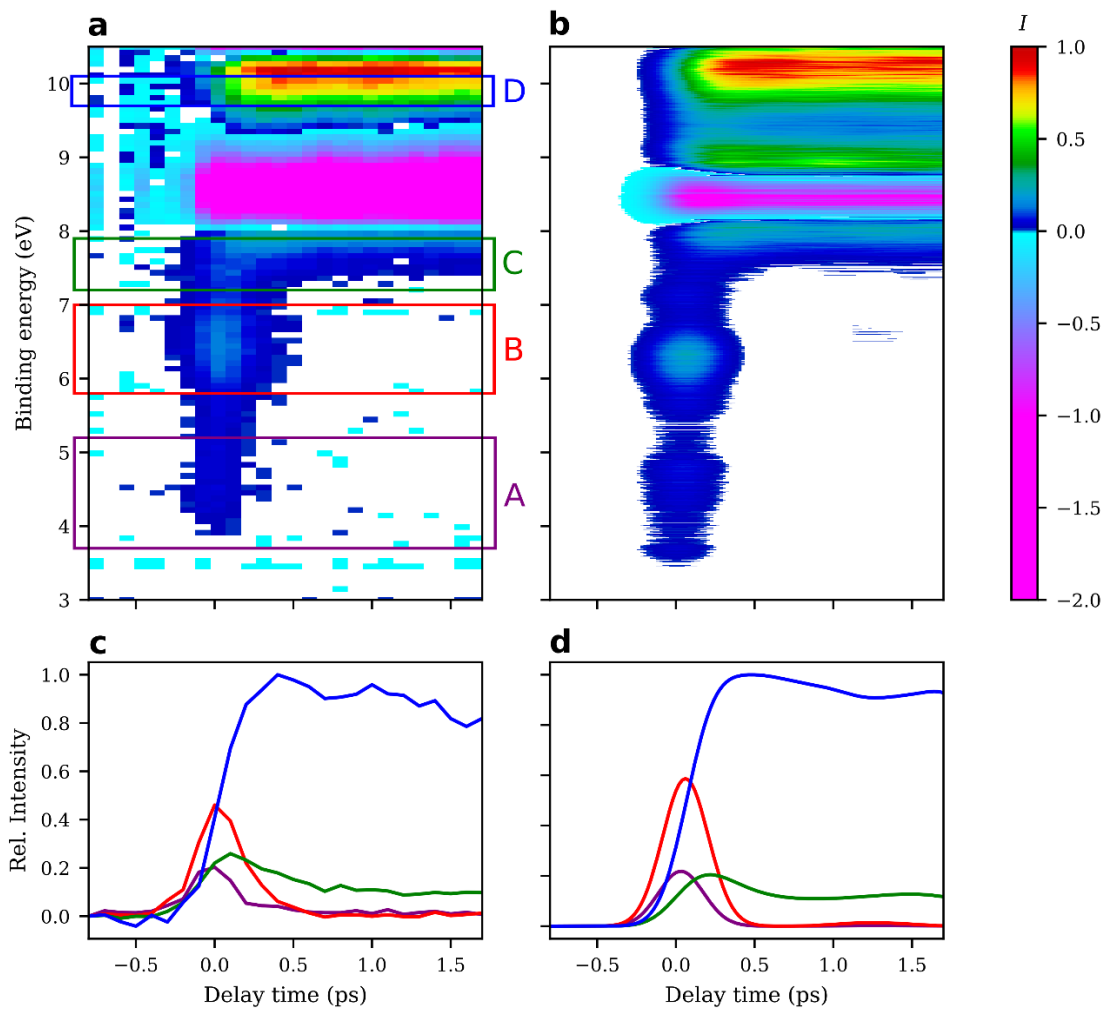

**Figure 1. Valence photoionization spectroscopy probes the time evolution of photoexcited CHD**. **a,** Experimental and **b**, Computed 2D maps of photoelectron spectra. Positive features correspond to an enhancement of the signal, the negative feature to a depletion. **c,** Experimental and **d**, Computed time evolution of the spectral intensity integrated over the area marked by colored rectangles in **(a)**. For experimental and computational details see the Supplementary Information.

The three features in the low-binding-energy region have a different behavior as a function of delay: the two peaks at 4.5 and 6.4 eV binding energy reach maxima at 0 ps and decrease to zero intensity after 0.5 ps, while the third feature at 7.5 eV has low intensity at 0 ps, continues to grow to a maximum at 0.5 ps and finally stabilizes after about 1 ps. The feature at high binding energy, 9.9 eV, grows after 0 ps as a function of delay, and stays constant after about 0.5 ps. We denote these features A, B, C and D in order of increasing binding energy. The disappearing features A and B are "feeding" the developing ones C and D, giving evidence of a structural change accompanying the electronic relaxation. This is a first hint that features A and B correspond to excited states which relax on a fs time scale, while the features C and D developing and then reaching an asymptotic value are connected to electronic states belonging to vibrationally hot open chain and/or closed ring isomers.

This assignment is strongly supported by theory. The photoinduced dynamics was simulated with non-adiabatic surface-hopping trajectories and the photoelectron spectra were computed from ionization energies and Dyson orbital norms. To match the experiment, the theoretical spectra have been shifted by +0.3 eV. For further details on how the theoretical spectra are calculated see the Methods section and SI. The agreement between the experimental and simulated spectra is remarkably good, both in terms of intensity and time evolution of the peaks, with the width of the ground-state bleach component (purple) being the only major discrepancy. The simulated spectra, when analyzed separately for non-reactive (CHD) and reactive (HT) trajectories (see Supplementary Fig. 5) show that band C arises mainly from vibrationally hot CHD, while the band D is dominantly due to newly formed HT molecules. The latter assignment is confirmed by a comparison between the ground-state valence photoelectron spectra of CHD and HT, where there is spectral intensity in the binding energy region around 10 eV only for the open-chain isomer [40,41].

To fully uncover the mechanism of the reaction, we need to connect the evolution of the photoionization bands as a function of time with the population of the diabatic electronic states involved in the reaction. To this end we first consider the properties of the electronic states at the Franck-Condon geometry.

The electronic spectrum of CHD reported in the literature is composed of two broad bands at ~5.0 and ~8.0 eV [42]. The first encompasses the bright $1^1A^-\to1^1B$ and dark $1^1A^-\to2^1A^-$ transitions. The second or cis band is characteristic for cis polyenes. Here two valence transitions of mixed character have been identified – the intense $1^1A^-\to1^1A^+$ transition at 8.0 eV [4,43] and a higher-lying transition with considerable vibronic structure [42]. Intercalated between these two bands are several sharp Rydberg transitions [4,43,44]. Owing to their weak coordinate dependence, Rydberg states are expected to play a negligible role in the ring-opening reaction.

Vertical excitation energies of the six lowest valence excited states of CHD with the coefficients of the leading configuration state functions (CSFs) are collected in Supplementary Table 1 (see also Supplementary Table 2), while the orbitals constituting the active space are illustrated in Supplementary Fig. 6. The main results are summarized in Figure 2. At the Franck-Condon geometry, the XMS(7)-CASPT2[6e,6o] method yields the $1^1B$ state at 5.18 eV, the dark $2^1A^-$ state at 5.99 eV and the two bright states of the *cis*-band, $1^1A^+$ and $2^1A^+$, at 8.31 and 8.66 eV, respectively. The electron density differences between the excited states and the ground state plotted in Fig. 2a show that excitation to the $1^1B$ and $2^1A^+$ states leads to depletion of electron density on the $C_1$-$C_6$ bond, while excitation to $2^1A^-$ and $1^1A^+$ leaves the density on that bond nearly unchanged. From the wave functions of the three states of symmetry A, which are dominated by the three CSFs shown in Fig. 2b, it is evident that the $2^1A^+$ state has a pronounced double-excitation character (see CI coefficients) and the overall electronic properties of a potentially reactive state.

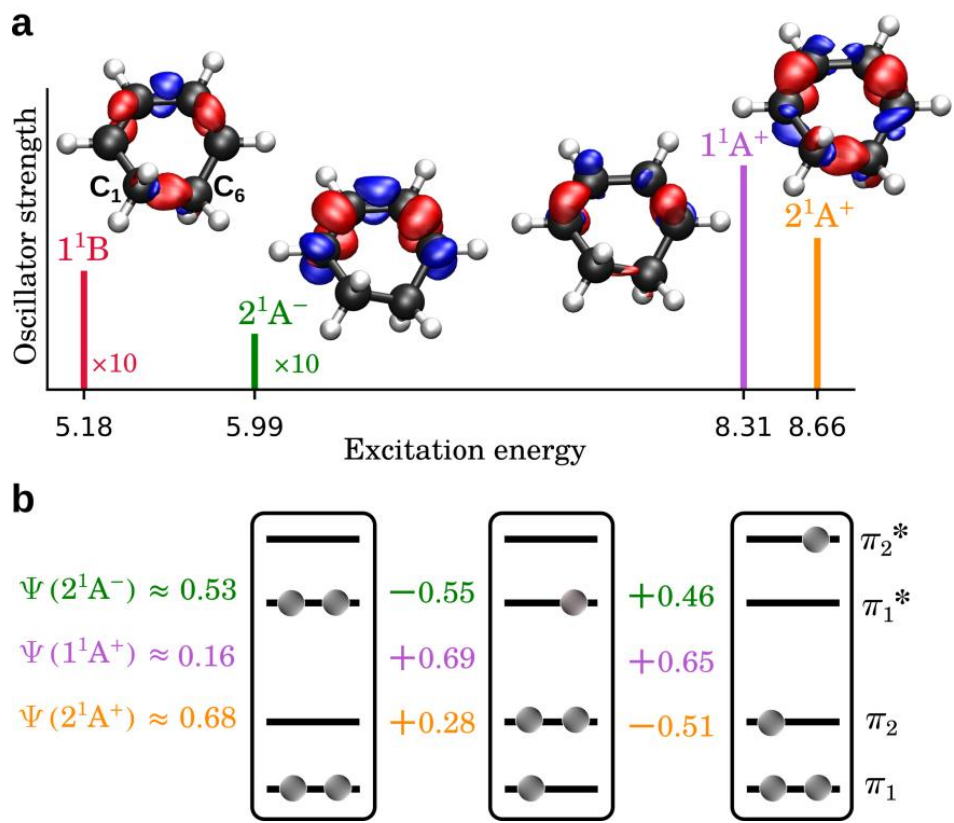

**Figure 2. The chemical character of the excited states at the Franck-Condon geometry explains their reactivity. a,** Excitation energies and relative absorption intensities of the $1^1B$, $2^1A^-$, $1^1A^+$ and $2^1A^+$ states (sticks) and the corresponding maps showing the electron density difference with respect to the electronic ground state calculated at the Franck-Condon geometry. Areas of increased and reduced electron density are shown in blue and red, respectively. **b,** The three valence states, $2^1A^-$ (green), $1^1A^+$ (purple) and $2^1A^+$ (orange), with the CI coefficients of the three dominant CSFs ($\pi_2\pi_2 \rightarrow \pi_1^*\pi_1^*$, $\pi_1 \rightarrow \pi_1^*$ and $\pi_2 \rightarrow \pi_2^*$).

Figure 3 shows the one-dimensional potential energy profiles of the lowest adiabatic (black) and diabatic (colours) states along the ring-opening path in $C_2$ symmetry. The diabatic potentials are obtained by minimizing the variation of their wave functions along the reaction path [45]. Details of the procedure are given in the SI. One immediately sees that the photochemical mechanism emerging from Figure 3 is not in agreement with the traditionally accepted one. While the deviation of the diabatic $1^1B$ state (red) first from the $S_1$ state and then from the $S_2$ state matches the expected behaviour, this is not the case for the $2^1A^-$ state (green) as the state clearly increases in energy (destabilizes) along the reaction path. The diabatic $2^1A^+$ state (orange) is the state that is strongly stabilized in the reaction. The two states cross at $R(C_1\text{-}C_6)$ ~1.9 Å where the adiabatic $S_2$ state shows the characteristic inflection. Further down the reaction path, $2^1A^+$ crosses with the $1^1B$ state and at $R(C_1\text{-}C_6)$ ~2.3 Å with the ground state. This indicates that the ground state of HT correlates with the $2^1A^+$ state and not with the $2^1A^-$ state. Notice that the proposed pathway does not contradict the Woodward-Hoffmann rules as they do not prescribe that the reaction should proceed on the lowest doubly excited state in the Franck-Condon geometry.

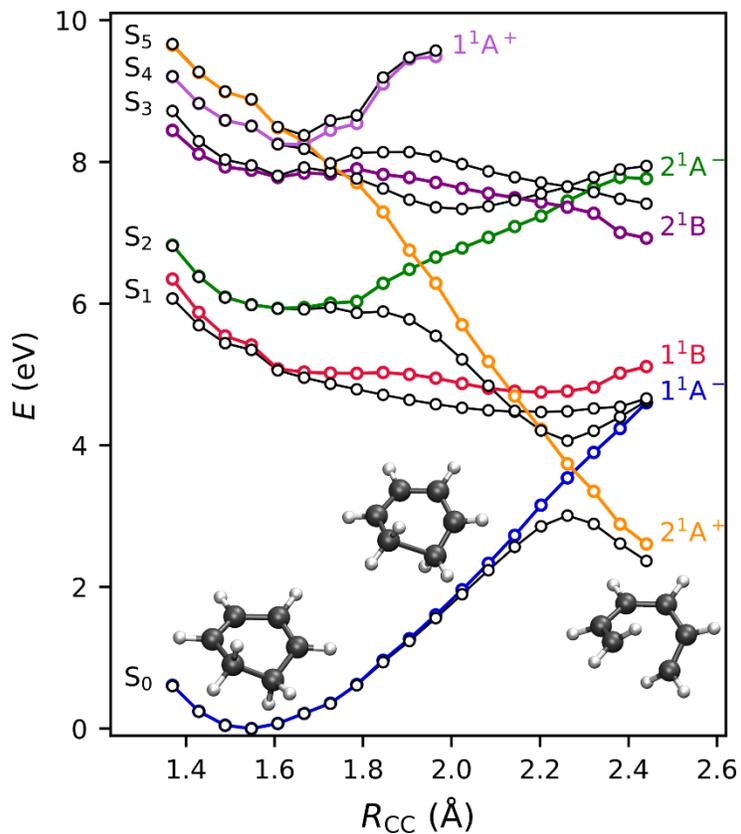

**Figure 3. Potential curves along the CHD ring opening path in $C_2$ symmetry reveal the strong stabilization of the $2^1A^+$ state.** Coordinate dependence of the potential energy of the lowest adiabatic electronic states (black) and diabatic electronic states: $1^1A^-$ (blue), $1^1B$ (red), $2^1A^-$ (green), $2^1B$ (dark violet), $1^1A^+$ (light violet) and $2^1A^+$ (orange). For details of the diabatization procedure see the Supporting Material.

To understand the role of the different diabatic states in the ring-opening reaction when the $C_2$ symmetry is lifted, we analyse two representative non-adiabatic trajectories yielding CHD (Fig 4a-4c) and HT (Fig. 4d-4f). For other trajectories see Supplementary Fig. 7. The two trajectories exhibit a very similar behaviour up to the CoIn with the ground state. In both cases the dynamics is initiated in the $S_1$ state (Fig 4a and 4d), which is in the diabatic $1^1B$ state (Fig. 4b and 4e). At two instances, at ~10 and ~25 fs, the gap between the $S_1$ and $S_2$ states becomes vanishingly small and the $S_1$ state has a possibility to change its character. Indeed, the diabatic populations show a brief increase of the $2^1A^-$ contribution at ~10 fs (green), but then the $1^1B$ character is recovered until ~25 fs when its contribution suddenly drops. From ~25 fs onward the $S_1$ state is best described as a superposition of the $1^1A^-$ (blue) and $2^1A^+$ (orange) diabatic states. The two states are strongly coupled by nuclear motion along the so-called extended bond alternation coordinate (BAC*), which is the difference of single and double bonds lengths in HT plus $R(C_1-C_6)$. [11] In the CHD trajectory the return to the ground state occurs after a BAC* local maximum, while the $C_1-C_6$ bond is compressing and the $S_1$ state is dominantly of $1^1A^-$ character. On the contrary, in the HT trajectory the CoIn with the ground state is encountered after a BAC* local minimum, while the $C_1-C_6$ bond is expanding and the $S_1$ state is dominantly of $2^1A^+$ character. After the hop the CHT trajectory continues to evolve in the $1^1A^-$ state and the HT trajectory in the $2^1A^+$ state, which is now the ground state leading to the HT product. Altogether, our analysis suggests that the fate of a nonadiabatic trajectory is determined by the character of the $S_1$ state at the moment of the $S_1 \rightarrow S_0$ nonadiabatic transition. If the transition occurs when the $S_1$ state has a dominant $1^1A^-$ character, the CHD product is formed and, *vice versa*, if the hop occurs when the $S_1$ state has a dominant $2^1A^+$ character the HT product is formed.

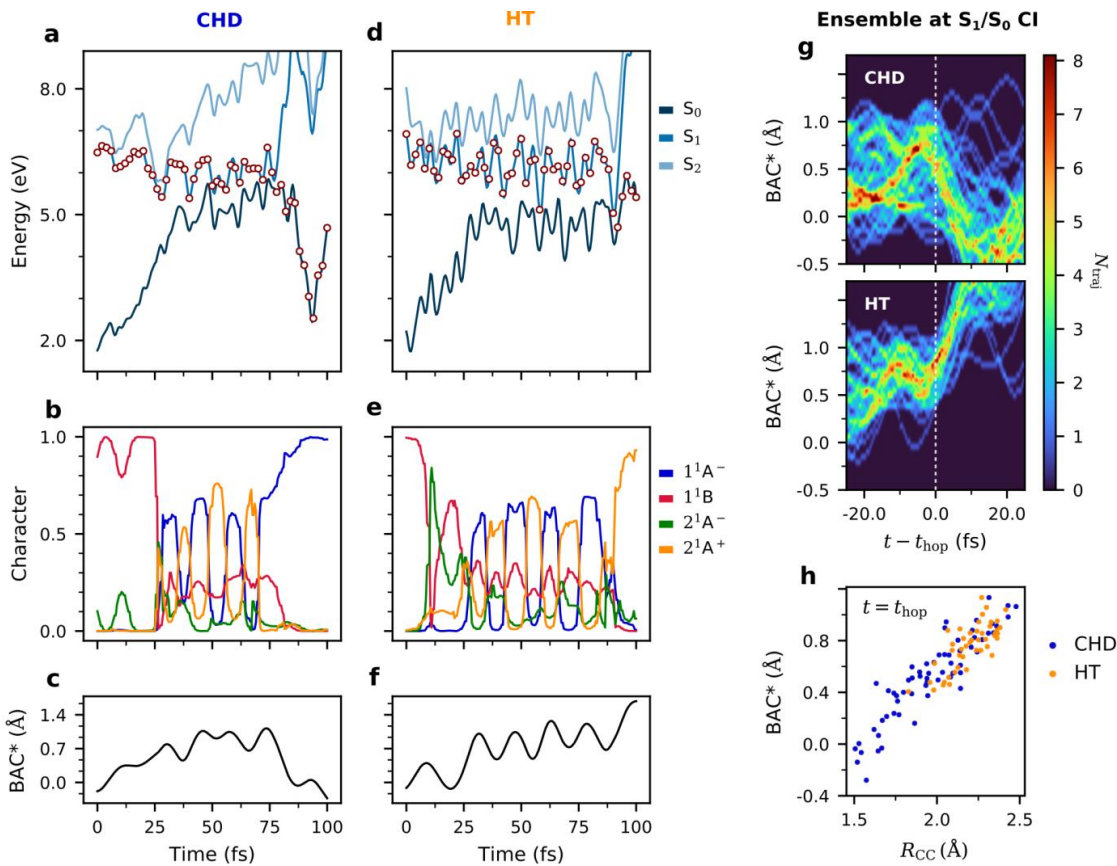

**Figure 4. Representative surface-hopping trajectories reveal that the fate of photoexcited CHD depends on the diabatic composition of the $S_1$ state at the moment of hop.**
**a)** and **d)** Representative nonadiabatic dynamics trajectories leading to CHD (**a**) and HT (**d**). Time evolution of the potential energy of the electronic ground state $S_0$ (dark blue) and the two lowest excited states $S_1$ (blue) and $S_2$ (light blue). Dots mark the instantaneous populations of electronic states. **b)** and **e)** Decomposition of the populated states in terms of four diabatic states, $1^1A^-$ (blue), $1^1B$ (red), $2^1A^-$ (green) and $2^1A^+$ (orange) along the trajectory. **c)** and **f)** Evolution of the extended bond alternation coordinate (BAC*) along the nonadiabatic trajectory. **g)** Evolution of the BAC* coordinate, which is the difference of the length of the single and double bonds in HT plus the length of the C1-C6 bond, for the ensemble of nonadiabatic trajectories synchronised to reach the $S_1/S_0$ CoIn simultaneously at $t' = t - t_{hop} = 0$. At the moment of hop BAC* is decreasing for CHD and increasing for HT trajectories. **h)** Distribution of $R(C_1\text{-}C_6)$ and BAC* at the moment of hop to $S_0$. Orange (blue) circles correspond to HT (CHD) trajectories.

The analysis of the ensemble of non-adiabatic trajectories, divided in two groups, CHD and HT, and synchronized in such a way as to reach the $S_1/S_0$ CoIn at the same time, is given in Figures 4g-4h. Fig. 4g shows that for all HT trajectories the BAC* is increasing before the hop to $S_0$, while it is decreasing for most but not all CHD trajectories. The distribution of $R(C_1-C_6)$ and BAC* at the time of hop (Fig. 4h) indicates that for large BAC* and $R(C_1-C_6)$ distances both CHD (blue) and HT (orange) can be formed in a close to 50:50 ratio but for small BAC* and short $R(C_1-C_6)$ only CHD is formed, irrespectively of whether BAC* is compressing or not. A closer inspection reveals that in this group of non-reactive trajectories the population of the $2^1A^+$ state is negligibly small (see Supplementary Fig. 8), meaning that the existence of a second non-reactive pathway from either $1^1B$ or $2^1A^-$ cannot be excluded.

We can now relate the average adiabatic and diabatic electronic populations to the time evolution of the two lowest-energy bands in the photoelectron spectra. Figure 5a shows the two-dimensional map of unconvoluted theoretical photoelectron spectra at short delay times. Two bands are clearly visible in the binding energy range of 3-7 eV. Band A starts at ~3.2 eV and within 15 fs reaches a plateau at ~4.5. eV. It arises from the $S_1 \rightarrow D_0$ transition. The increase of the ionization energy is caused by the motion toward the minimum of the $S_1$ state. As this motion leads to the extension of the $C_1-C_6$ bond, the energy of the ground state of the CHD cation ($D_0$) increases (see Supplementary Fig. 9). At around ~30 fs the band loses intensity.

Band B starts at ~7.0 eV but it is almost immediately stabilized to the 5.8-6.2 eV binding energy range. The sudden stabilization is caused by the crossing of the cationic $D_1$ and $D_2$ states (see Supplementary Fig. 9). The maximum of the intensity of band B is reached at ~40 fs. Fig. 5b shows the average population of the adiabatic states $S_0$, $S_1$ and $S_2$ as obtained from nonadiabatic dynamics simulations. By comparing the time evolution of bands A and B with the average population of the adiabatic states one sees that the loss of intensity of band A coincides with the transfer of population from the $S_1$ to the $S_2$ state with maximum at ~30 fs, while the maximum of the intensity of band B at ~40 fs coincides with the rather counterintuitive rise of the population of the $S_0$ state. The electronic populations of the diabatic states (Fig. 5c) provide a more consistent view. The population initially residing in the $1^1B$ state steadily decreases. The $2^1A^-$ state is transiently populated at early times but its population never exceeds 0.2. In the ~35-55 fs interval the population of the $1^1A^-$ (blue) and $2^1A^+$ (orange) states increases and the system evolves in a superposition of these two states. This interval coincides with the maximum of the intensity of band B. The decomposition of the photoelectron spectra into contributions from diabatic states [46] in Figs. 5d-5g unambiguously shows that the increase of the intensity of band B at around 40 fs originates from the population of the $2^1A^+$ state (Fig. 5g). As the reaction proceeds, the formation of the CHD and HT product correlating with the $1^1A^-$ and $2^1A^+$ states, respectively, becomes clearly visible.

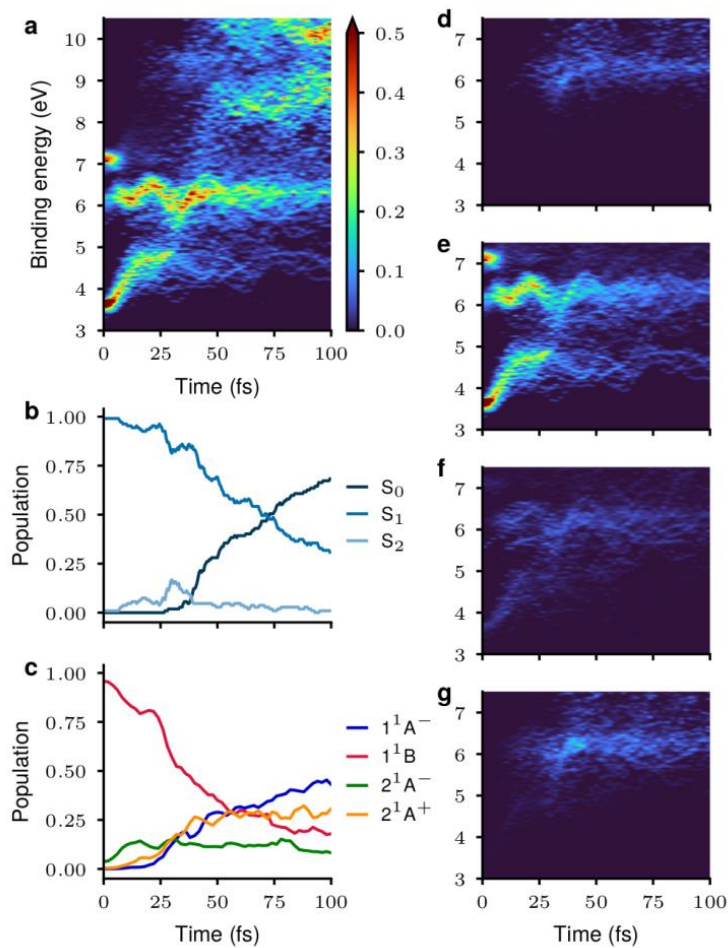

**Figure 5**. **Analysis of the photoelectron spectra in terms of electronic populations of diabatic states. a)** Unconvoluted photoelectron spectra for short delay times. The ground state bleach component is not taken into account. **b)** Time evolution of the adiabatic population of electronic states obtained from surface hopping nonadiabatic dynamics simulations. **c)** Time evolution of the diabatic population of electronic states obtained by diabatization of 107 non-adiabatic trajectories. **d-f**, Decomposition of the photoionization spectrum in terms of contributions of the diabatic states $1^1A^-$ (d), $1^1B$ (e), $2^1A^-$ (f), $2^1A^+$ (g).

Our analysis clearly points towards a predominant role of the $2^1A^+$ diabatic state in the reactive path, and in general a more consistent picture of the overall dynamics can be achieved by analyzing diabatic rather than adiabatic contributions. We consider our findings as a step forward with respect to the three-step description usually reported in the literature [11-19]. Although our experimental results do not contradict the previous ones, here we show the limits of the usually accepted model. We rely on both experimental advances, such as the possibility of measuring time-resolved valence photoelectron spectra with rather high electron energy resolution, and theoretical methods allowing accurate calculation of the photoelectron signal and on-the-fly adiabatic-to-diabatic transformation of electronic populations.

## Conclusion

The photochemical ring-opening reaction of 1,3-cyclohexadiene (CHD) to 1,3,5-hexatriene (HT) is a textbook example of a pericyclic reaction, and possibly the most investigated isomerization reaction with advanced time-resolved spectroscopies. Here we show that the usual description of the reaction pathway in three steps (photoexcitation to the lowest-lying bright state followed by a conical intersection with the first dark state and then either another conical intersection with the open-chain isomer ground state or a return to the cyclic isomer ground state) is oversimplified, and the overall picture is much more consistent if diabatic rather than adiabatic states are analyzed. In particular, we show that the doubly excited dark state, labeled $2^1A^-$, which is considered in the literature the gateway to the isomerization process, does not play a significant role. Instead, an initially high-lying state, labeled $2^1A^+$, with a pronounced double excitation character and a significant reduction of electron density upon the $C_1$-$C_6$ bond (the one which breaks during the isomerization process), is the reactive state whose temporal evolution drives the reaction. We consider our results as a general conceptual advance in the field of ultrafast dynamical studies by time-resolved spectroscopies.

## Acknowledgment

We are grateful to the FERMI Team and laser group, in particular to L. Giannessi, M. B. Danailov, and A. Demidovich, for their continuous support during the experiments. T.P, A.P, M.Sa, P.D. and N. D thank the Croatian Science Foundation for financial support (IP-2016-06-1142 and IP-2020-02-9932). R.F. thanks the Swedish Research Council and the Knut and Alice Wallenberg Foundation for financial support. We acknowledge the support of the COST Action CA18222 (Attosecond Chemistry). Many fruitful discussions with J.H.D. Eland and W. Domcke are gratefully acknowledged.

## Author contribution

M.N.P., R.F, K.C.P. and M.Si. devised the research, O. T., R.J.S., R.R., P.F., M. Di F., A. De F., N. M., M. M., V.Z., T.M., R.G., L. J., C.C., K.C.P., M.Si., R.F., and M.N.P. participated in the conduction of the experimental research, O. T. performed the data analysis, T. P., A. P., M.Sa., P.D., and N.D. devised the theoretical procedure and performed the calculations, M.N.P. and N.D. wrote the paper and all authors discussed the results and commented on the manuscript.

## Data availability

The data sets generated during and/or analyzed during the current study are available upon reasonable request from the corresponding authors (M.N.P. and N.D.).

## Competing interests

The authors declare no competing financial interests.

# Methods

*Experimental*

The experiments were performed at the low-density matter (LDM) beamline [37] at the FERMI free-electron laser facility. The Ti:Sapphire optical laser (pump) was operated at 267 nm, with a bandwidth of 1.2 nm. The FEL pulse (probe) was set at a photon energy of 19.23 eV, corresponding to the fourth harmonic of the seed wavelength of 258 nm. The spectrometer used to collect photoelectrons was a magnetic bottle [38,39].

Electron time-of-flight (TOF) spectra were recorded shot-by-shot while the delay between the pump and probe pulses was scanned with a step of 100 fs. The data used to construct the Fig. 1 consist of 15000 shots per each delay, which were summed and normalized by the summed FEL intensity, recorded simultaneously for every shot. The electron flight times were converted to electron kinetic energies and calibrated according to Ref. [40] with respect to the FEL photon energy. The influence of the FEL and UV intensity on the TOF spectral shape was verified by performing a set of measurements at varied pulse energies and UV focus values.

See SI for further details of experimental parameters, sample handling and data analysis.

*Theory*

In all calculations, CHD and HT were described by extended multi-state complete active space self-consistent field second-order perturbation theory (XMS-CASPT2) [47,48] employing an active space of 6 electrons in 6 orbitals, CAS(6e,6o) (see Supplementary Fig. 6). In nonadiabatic dynamics simulations, the CASSCF orbitals were averaged over three states with equal weights. The three lowest electronic states of the CHD/HT cation ($D_0$, $D_1$, $D_2$) were taken into account in the computation of the photoelectron spectra. The cc-pVDZ basis set was used in all computations and a real shift of 0.5 $E_h$ was employed to avoid intruder states in the dynamics. All electronic-structure calculations were performed with the BAGEL program [49,50].

The photoelectron spectra were computed using the classical limit of the doorway-window formalism [51] Nonadiabatic dynamics simulations were performed with the Tully's fewest switches surface hopping (FSSH) [52] algorithm using an in-house code. The initial conditions were selected from the classical doorway function describing the excitation of the system by the pump pulse in our experiment.

Diabatic states were obtained from the adiabatic states by employing the diabatization scheme of Simah, Hartke, and Werner [45]. To obtain smooth diabatic potentials we included seven states in the calculations (XMS(7)-CASPT2[6e,6o]). Details of the nonadiabatic dynamics simulations, computation of photoelectron spectra and the diabatization procedure are given in the Supplementary Information.

# Literature references

[1] Deb, S. & Weber, P. M. The ultrafast pathway of photon-induced electrocyclic ring-opening reactions: the case of 1,3-cyclohexadiene. *Annu. Rev. Phys. Chem.* **62,** 19–39 (2011).

[2] Arruda, B. C. & Sension, R. J. Ultrafast polyene dynamics: the ring opening of 1,3-cyclohexadiene derivatives. *Phys. Chem. Chem. Phys.* **16,** 4439 (2014).

[3] Celani, P., OIivucci, M., Bernard, F., Ottani, S. & Robb, M.A. What happens during the picosecond lifetime of 2A1 cyclohexa-1,3-diene? A CAS-SCF study of the cyclohexadiene/hexatriene photochemical interconversion. *J. Am. Chem. Soc.* **116,** 10141–10151 (1994).

[4] Merchán, M. *et al*. Electronic spectra of 1,4-cyclohexadiene and 1,3-cyclohexadiene: a combined experimental and theoretical investigation. *J. Phys. Chem. A* **103,** 5468–5476 (1999).

[5] Garavelli, M. *et al*. Reaction path of a sub-200 fs photochemical electrocyclic reaction. *J. Phys. Chem. A* **105,** 4458–4469 (2001).

[6] Nenov, A., Kölle, P., Robb, M.A. & de Vivie-Riedle, R. Beyond the van der Lugt/Oosterhoff model: When the conical intersection seam and the $S_1$ minimum energy path do not cross. *J. Org. Chem.* **75,** 123–129 (2010).

[7] Hofmann, A. & de Vivie-Riedle, R. Quantum dynamics of photoexcited cyclohexadiene introducing reactive coordinates. *J. Chem. Phys.* **112,** 5054–5059 (2000).

[8] Schönborn, J.B., Sielk, J. & Hartke, B. Photochemical ring-opening of cyclohexadiene: Quantum wavepacket dynamics on a global ab initio potential energy surface. *J. Phys. Chem. A* **114,** 4036–4044 (2010).

[9] Ohta, A., Kobayashi, O., Danielache, S.O. & Nanbu, S. Nonadiabatic ab initio molecular dynamics of photoisomerization reaction between 1,3-cyclohexadiene and 1,3,5-cis-hexatriene. *Chem. Phys.* **459,** 45–53 (2015).

[10] Lei, Y., Wu, H., Zheng, X., Zhai, G. & Zhu, C. Photo-induced 1,3-cyclohexadiene ring opening reaction: Ab initio on-the-fly nonadiabatic molecular dynamics simulation. *J. Photochem. Photobiol. A Chem.* **317,** 39–49 (2016).

[11] Schalk, O. *et al*. Cyclohexadiene revisited: A time-resolved photoelectron spectroscopy and ab initio study. *J. Phys. Chem. A* **120,** 2320–2329 (2016).

[12] Polyak, I., Hutton, L., Crespo-Otero, R., Barbatti, M. & Knowles, P.J. Ultrafast photoinduced dynamics of 1,3-cyclohexadiene using XMS-CASPT2 surface hopping. *J. Chem. Theory Comput.* **15,** 3929–3940 (2019).

[13] Kotur, M., Weinacht, T., Pearson, B.J. & Matsika, S. Closed-loop learning control of isomerization using shaped ultrafast laser pulses in the deep ultraviolet. *J. Chem. Phys.* **130,** 134311 (2009).

[14] Attar, A.R. *et al*. Femtosecond x-ray spectroscopy of an electrocyclic ring-opening reaction. *Science* **356,** 54–59 (2017).

[15] Adachi, S., Sato, M. & Suzuki, T. Direct observation of ground-state product formation in a 1,3-cyclohexadiene ring-opening reaction. *J. Phys. Chem. Lett.* **6,** 343–346 (2015).

[16] Pemberton, C.C., Zhang, Y., Saita, K., Kirrander, A. & Weber, P.M. From the (1B) spectroscopic state to the photochemical product of the ultrafast ring-opening of 1,3-cyclohexadiene: A spectral observation of the complete reaction path. *J. Phys. Chem. A* **119,** 8832–8845 (2015).

[17] Petrović, V.S. *et al*. Transient X-ray fragmentation: probing a prototypical photoinduced ring opening. *Phys. Rev. Lett.* **108,** 253006 (2012).

[18] Wolf, T.J.A. *et al*. The photochemical ring-opening of 1,3-cyclohexadiene imaged by ultrafast electron diffraction. *Nat. Chem.* **11,** 504–509 (2019).

[19] Karashima, S. *et al*. Ultrafast ring-opening reaction of 1,3-cyclohexadiene: identification of nonadiabatic pathway via doubly excited state. *J. Am. Chem. Soc.* **143,** 8034–8045 (2021).

[20] Havinga, E. & Schlatmann, J.L.M.A. Remarks on the specificities of the photochemical and thermal transformations in the vitamin D field. *Tetrahedron* **16,** 146–152 (1961).

[21] Anderson, N.A., Shiang, J.J. & Sension, R.J. Subpicosecond ring opening of 7-dehydrocholesterol studied by ultrafast spectroscopy. *J. Phys. Chem. A* **103,** 10730–10736 (1999).

[22] Matsuda, K. & Irie, M. Diarylethene as a photoswitching unit. *J. Photochem. Photobiol. C Photochem. Rev.* **5,** 169–182 (2004).

[23] Kobatake, S., Takami, S., Muto, H., Ishikawa, T. & Irie, M. Rapid and reversible shape changes of molecular crystals on photoirradiation. *Nature* **446,** 778–781 (2007).

[24] Irie, M., Fukaminato, T., Matsuda, K. & Kobatake, S. Photochromism of diarylethene molecules and crystals: memories, switches, and actuators. *Chem. Rev.* **114,** 12174–12277 (2014).

[25] Dattler, D. *et al*. Design of collective motions from synthetic molecular switches, rotors, and motors. *Chem. Rev.* **120,** 310–433 (2020).

[26] Baroncini, M., Silvi, S. & Credi, A. Photo- and redox-driven artificial molecular motors. *Chem. Rev.* **120,** 200–268 (2020).

[27] Domcke, W., Yarkony, D.R. & Köppel, H. *Conical Intersections: Theory, Computation and Experiment* (World Scientific, Singapore, 2011).

[28] Nakamura, H. *Nonadiabatic Transition: Concepts, Basic Theories and Applications* (World Scientific, Singapore, 2012).

[29] Yarkony, D.R. Nonadiabatic quantum chemistry—past, present, and future. *Chem. Rev.* **112,** 481–498 (2012).

[30] Domcke, W. & Stock, G. Theory of ultrafast nonadiabatic excited-state processes and their spectroscopic detection in real time, *Adv. Chem. Phys.* **100**, 1-169 (1997)

[31] Woodward, R.B. & Hoffmann, R. The conservation of orbital symmetry. *Angew. Chemie Int. Ed.* **8,** 781–853 (1969).

[32] Van der Lugt, W.T.A.M & Oosterhoff, L.J. Quantum-chemical interpretation of photo-induced electrocyclic reactions. *Chem. Commun.* 1235 (1968).

[33] Van der Lugt, W.T.A.M. & Oosterhoff, L.J. Symmetry control and photoinduced reactions. *J. Am. Chem. Soc.* **91,** 6042–6049 (1969).

[34] Pariser, R. Theory of the electronic spectra and structure of the polyacenes and of alternant hydrocarbons. *J. Chem. Phys.* **24,** 250–268 (1956).

[35] Squibb, R.J. *et al*. Acetylacetone photodynamics at a seeded free electron laser. *Nat. Commun.* **9,** 63 (2018).

[36] Pathak, S. *et al*. Tracking the ultraviolet-induced photochemistry of thiophenone during and after ultrafast ring opening. *Nat. Chem.* **12,** 795–800 (2020).

[37] Svetina, C. *et al*. The low density matter (LDM) beamline at FERMI: optical layout and first commissioning. *J. Synchrotron Radiat.* **22,** 538–543 (2015).

[38] Eland, J.H.D. *et al*. Complete two-electron spectra in double photoionization: the rare gases Ar, Kr, and Xe. *Phys. Rev. Lett.* **90,** 053003 (2003).

[39] Eland, J.H.D., Linusson, P., Mucke, M. & Feifel, R. Homonuclear site-specific photochemistry by an ion–electron multi-coincidence spectroscopy technique. *Chem. Phys. Lett.* **548,** 90–94 (2012).

[40] Kimura, K. *et al*., Handbook of HeI Photoelectron Spectra of Fundamental Organic Molecules, Japan Scientific Societies Press, Tokyo, p. 68.

[41] Beez, M., Bieri, G., Bock, H. & E. Heilbronner, Helvetica Chim. Acta 56 (1973) 1028

[42] Robin, M.B. Higher Excited States of Polyatomic Molecules Vol. 2 (Academic, New York, 1975).

[43] McDiarmid, R., Sabljić, A. & Doering, J.P. Valence transitions in 1,3- cyclopentadiene, 1,3- cyclohexadiene, and 1,3- cycloheptadiene. *J. Chem. Phys.* **83,** 2147–2152 (1985).

[44] Ning, J. & Truhlar, D.G. The valence and Rydberg states of dienes. Phys. Chem. Chem. Phys. 22, 6176–6183 (2020).

[45] Simah, D., Hartke, B. & Werner, H.-J. Photodissociation dynamics of H2S on new coupled ab initio potential energy surfaces. *J. Chem. Phys.* **111,** 4523–4534 (1999).

[46] Piteša, T. *et al*. Combined Surface-Hopping, Dyson Orbital, and B-Spline Approach for the Computation of Time-Resolved Photoelectron Spectroscopy Signals: The Internal Conversion in Pyrazine. *J. Chem. Theory Comput.* **17,** 5098–5109 (2021).

[47] Shiozaki, T., Győrffy, W., Celani, P. & Werner, H.-J. Communication: Extended multi-state complete active space second-order perturbation theory: Energy and nuclear gradients. *J. Chem. Phys.* **135,** 081106 (2011).

[48] Park, J.W. & Shiozaki, T. On-the-fly CASPT2 surface-hopping dynamics. *J. Chem. Theory Comput.* **13,** 3676–3683 (2017).

[49] Shiozaki, T. BAGEL: Brilliantly advanced general electronicstructure library. Wiley Interdiscip. Rev.: Comput. Mol. Sci. 8, el1311 (2018).

[50] Shiozaki, T. BAGEL: Brilliantly Advanced General Electronic-structure Library. http://www.nubakery.org under the GNU General Public License

[51] Gelin, M.F. *et al*. Ab initio surface-hopping simulation of femtosecond transient-absorption pump–probe signals of nonadiabatic excited-state dynamics using the doorway–window representation. *J. Chem. Theory Comput.* **17,** 2394–2408 (2021).

[52] Tully, J. Molecular dynamics with electronic transitions. *J. Chem. Phys.* **93,** 1061 (1990).

# Supplementary information

for

# The photochemical ring-opening reaction of 1,3-cyclohexadiene: complex dynamical evolution of the reactive state

O. Travnikova[1,#], T. Piteša[2,#], A. Ponzi[2], M. Sapunar[2], R. J. Squibb[3], R. Richter[4], P. Finetti[4], M. Di Fraia[4], A. De Fanis[5], N. Mahne[6], M. Manfredda[4], V. Zhaunerchyk[3], T. Marchenko[1], R. Guillemin[1], L. Journel[1], K. C. Prince[4], C. Callegari[4], M. Simon[1], R. Feifel[3], P. Decleva[7], N. Došlić[2,*] and M.N. Piancastelli[1,8,*]

[1]*Sorbonne Université, CNRS, Laboratoire de Chimie Physique-Matière et Rayonnement, LCPMR, F-75005, Paris, France*

[2]*Institut Ruđer Bošković, Bijenička cesta 54, HR-10000 Zagreb, Croatia*

[3]*Department of Physics, University of Gothenburg, Origovägen 6B, SE-412 96 Gothenburg, Sweden*

[4]*Elettra-Sincrotrone Trieste, Strada Statale 14-km 163.5, 34149 Basovizza, Trieste, Italy*

[5]*European XFEL, D-22869 Schenefeld, Germany.*

[6]*IOM-CNR, S.S. 14 km 163.5 in Area Science Park, 34149 Trieste (Italy)*

[7]*Dipartimento di Scienze Chimiche e Farmaceutiche, Universitá di Trieste, I-34127 Trieste, Italy.*

[8]*Department of Physics and Astronomy, Uppsala University, SE-751 20 Uppsala, Sweden*

[#] These authors have contributed equally.

*To whom correspondence should be addressed, maria-novella.piancastelli@physics.uu.se, nadja.doslic@irb.hr

# Table of content

# 1. Experimental

## 1.1. Beam parameters

The beam profile of the pump laser, SLU, (wavelength: 266 nm, bandwidth 1.2 nm FWHM, duration 120-130 fs), was set to a circular spot of ~500 x 500 μm FWHM, so as to completely illuminate the profile of the FEL beam; the latter is adjusted by mechanically changing the shape of the focusing Kirkpatrick-Baez (K-B) mirrors [1]. At these long wavelengths, the divergence of the incoming FEL beam is large, and the vignetting due to the photon transport optics [1] results in a rectangular profile; the shape of the K-B was set for a profile of 300 μm x 300 μm. The polarization of both beams was set to horizontal.

The spatial and temporal overlap ($t_0$) of the FEL and SLU pulses were optimized by setting the FEL undulators to the 5th harmonic of the seed laser (257.85 nm), corresponding to 51.57 nm, i.e., the 1s→5p resonance of the neutral He atom. The He+ ion yield of the resonant two-color two-photon (1+1') ionization was used as target signal for the manual optimization. The target spatial overlap is maintained over the course of the experiment with a precision of a few μm by means of an active feedback system, which tracks the position of a virtual beam on a commercial camera. Let us note that the pump and FEL pulses are transported along different paths, in a quasi-collinear arrangement [2].

For the specific experimental conditions, the FEL harmonic was then set to the 4th harmonic of the same seed wavelength (19.23 eV), to avoid ionizing the helium carrier gas. Higher harmonics of the fundamental undulator resonance (in this case: the 8th and 12th harmonic of the seed) are normally present at the level of 0.1–1% and were abated by a suitable pressure of Ne in the gas attenuator that is part of the photon transport system [3]; let us note that Ne is transparent at the fundamental wavelength. Because the attenuator was then unavailable for its primary purpose, the FEL intensity was instead varied machine-side, via a combination of the strength of the seeding process and the number of active undulator modules. The pump laser is attenuated with a combination of rotatable and fixed polarizers. The pulse energy of the pump laser is measured with a commercial energy meter ((Gentech S-Link; Head QE8SP-B-BL-D0), and corrected to account for measured transport losses; the pulse energy of the FEL is measured with a gas energy meter [4], and corrected for the calculated transport losses [1].

## 1.2. Spectrometer

The electron kinetic energy spectra and photo-ion spectra were measured using a tandem electron-ion time-of flight spectrometer based on a modified magnetic bottle electron spectrometer (MBES) [5,6]. The MBES operates as follows: by using a combination of a strong magnetic field located at the ionisation region, and a weak, homogenous magnetic field directed along the flight tube, photoelectrons which are initially emitted in all directions will have their trajectories parallelised towards the electron detector. The MBES therefore allows electrons to be collected over the full 4 π steradian solid angle and be detected with high efficiency (50-60% when microchannel plate efficiencies are included).
The weak magnetic field is generated by a solenoid with turn density 500/m, with a current of 0.3-2.0 ampère generating a magnetic field of the order of a few milliTesla. The solenoid starts ~ 90 mm from the interaction region and along the entire length of the ~2 m flight tube. A series of electrostatic lenses are positioned between the interaction region and the entrance to the flight tube in order to retard the electrons' kinetic energies, thus enhancing the resolution. The strong magnetic field is produced by three NeFeB hollow cylindrical permanent magnets with a conical soft iron polecap used to concentrate the magnetic field. The magnet assembly produces a peak field strength of ~ 0.1 Tesla, and yields a dE/E of approx. 1/20 – 1/30.

Additionally, the hollow profile of the magnet assembly allows ions that are produced by the FEL to be accelerated through the magnet by two additional electrostatic lenses and subsequently a flight tube, with a total interaction region to detector distance of 200-205 mm and a total ion acceleration of 400 V to 5 kV depending on the experimental conditions. The electrons and ions are each detected by two Hamamatsu F9892-31 chevron microchannel plate assemblies located at the far ends of the two flight tubes. The electron and ion signals are sent to separate channels of a CAEN analogue to digital system and the full waveforms stored for later analysis.

The spectrometer can be operated in either a pulsed mode or a static mode whereby low, DC voltages are used to extract and accelerate the electrons. This also causes the electrons to be accelerated, however the lenses located just before the electron flight tube can be used to retard the electrons to their original kinetic energy or less to maintain the electron resolution. For higher masses, where higher extraction fields are required, the voltage applied between the magnet and the first electron lens element is initially kept at close to ground to allow the electrons to escape the interaction region. After a 1-200 ns delay, a Behlke GHTS solid-state switch is used to apply a voltage to extract the ions.

### 1.3. Sample handling

The sample was mixed with 1.8 bar He in an external stainless steel bubbler, and injected into the experimental chamber via a commercial pulsed valve (Parker, Model 9, nominal aperture diameter: 800 µm). The bubbler was kept in a temperature-controlled refrigerated bath, and the partial pressure of the sample molecule was determined by the bath temperature under the assumption of thermodynamic equilibrium; the literature values used were taken from [7]. The valve was operated at the repetition rate of the FEL (50 Hz); the relative delay between the valve opening and the FEL pulse was scanned to determine sample-pulse length (~250 µs FWHM, depending on the nominal opening time) and optimal synchronization conditions. If desired, one out of n shots can be a blank shot (specifically: one for which the sample valve is fired out-of-sync with the FEL), and the measurement subtracted from the regular shots, after proper scaling. This is advantageous to assess the contribution of residual gas in the chamber; where applicable, we report the number n=6 as "background period". Changing the delay of a blank shot is preferable to suppressing it altogether, to preserve the thermal stability of the valve.

### 1.4. Data analysis

Photoelectron time-of-flight (TOF) traces were measured shot-by-shot for different delay times between pump and probe pulses. The FEL pulse energy was recorded for every shot and its distribution is shown in Fig. S1. The shots with very low (below 3 µJ) and high (above 17 µJ) FEL intensity were excluded from the data analysis. The TOF traces were summed and normalised to the summed FEL pulse intensity after subtraction of the «blank» background shots. Then the photoelectron flight times were transformed to electron kinetic energies by a non-linear conversion, which was derived by correlation of the peak maxima to the literature values from Ref.[8] subtracted from the FEL photon energy (19.23 eV). The electron kinetic energies scale was then converted to the binding energy scale. In total 15000 shots were recorded per each delay point for the pump-probe delay time scans, which were recorded with the step of 100 fs. Photoelectron spectra recorded for the time delay $t_0$, corresponding to the overlap of the UV-pump and FEL-probe pulses, UV pulse only and FEL pulse only are shown in Fig.S2. Evolution of the photoelectron spectra as function of the pump-probe delay is presented in Fig. S3 without subtraction of the ground-state spectrum. In Fig. 1 of the main manuscript, the ground-state spectrum was subtracted without adjustment of the spectral intensities. For this, the spectrum recorded at the negative delay time t=-0.7 ps was subtracted from all the photoemission spectra recorded at different delays. At t=-0.7 ps the FEL pulse arrives before the UV pulse, therefore only the ground-state CHD contributes to the photoemission.

Reference photoelectron spectra with only the FEL or UV pulse were recorded by accumulating in total 30000 shots per spectrum and a photoelectron spectrum when the FEL and UV pulses are overlapped at $t_0$ (pump-probe delay time t=0) was recorded by accumulating about 50000 shots. The zoom of these reference spectra is presented in Figure S3 for the electron binding energy region of interest.

An additional analysis was performed to check the quality of accumulated datasets, where the spread of the signal intensity during long pump-probe delay scans was plotted for each delay point. To do this, the traces were normalised to the FEL pulse intensity on shot-by-shot basis, then binned by 100 and the intensities, integrated over the specified TOF range, were histogrammed for each pump-probe delay point. The TOF regions, selected for this analysis, correspond to the binding energy regions, where the changes in the course of the ring-opening reaction of CHD are observed. The dataset with the least spread of the signal intensities was used for the constructions of Fig. 1 of the main manuscript and Fig. S4. The median distibutions of the intensities, integrated over the TOF regions of interest, were obtained as the center of gravity of the plotted histograms. These median distributions are qualitativitly similar to the integrated areas of the 2D photoelectron map presented in Fig. 1c of the main paper and allow tracking evolution of the photoelectron signal as a function of the pump-probe delay.

## 1.5. References

[1] Svetina, C. *et al*. The low density matter (LDM) beamline at FERMI: optical layout and first commissioning. J. Synchrotron Radiat. 22, 538–543 (2015).
[2] Finetti, P et al. Optical setup for two-colour experiments at the low density matter beamline of FERMI. *J. Opt.* **19,** 114010 (2017).
[3] Zangrando, M. *et al*. The photon analysis, delivery, and reduction system at the FERMI@Elettra free electron laser user facility. *Rev. Sci. Instrum.* **80,** 113110 (2009).
[4] Zangrando, M. *et al.* Recent results of PADReS, the photon analysis delivery and reduction system, from the FERMI FEL commissioning and user operations. *J. Synchrotron Rad.* **22,** 565–570 (2015).
[5] Kruit, P. & Read, F.H. Magnetic field paralleliser for $2\pi$ electron-spectrometer and electron-image magnifier. *J. Phys. E: Sci. Instrum.* **16,** 313 (1983).
[6] Eland, J.H.D. & Feifel, R. Double ionisation of ICN and BrCN studied by a new photoelectron–photoion coincidence technique. *Chem. Phys.* **327,** 85-90 (2006).
[7] https://webbook.nist.gov/cgi/cbook.cgi?ID=C592574
[8] K.Kimura *et al*., Handbook of HeI Photoelectron Spectra of Fundamental Organic Molecules, Japan Scientific Societies Press, Tokyo, pg. 68],

## 2. Computational

### 2.1. Static calculations

All electronic structure calculations were performed using extended multi-state (XMS) complete active space self-consistent field second-order perturbation (XMS-CASPT2) theory [1–3]. For the neutral CHD an active space of 6 electrons in 6 orbitals was used (CAS(6,6)), while an active space of 5 electrons in 6 orbitals (CAS(5,6)) was used for the cation. The orbitals constituting the active space computed at the equilibrium geometry of CHD are shown in the Supplementary Fig. 6. Nonadiabatic dynamics simulations were carried out in the manifold of the three lowest electronic states ($S_0$, $S_1$, $S_2$) of neutral CHD (XMS(3)-CASPT2(6,6)). A real shift of 0.5 $E_h$ was used to avoid intruder states in the dynamics. To compute the photoionization spectrum the three lowest electronic states of the CHD cation ($D_0$, $D_1$, $D_2$) were taken into account. The cc-pVDZ basis set was employed in all computations.

For the computation of the diabatic states along the ring opening reaction path and along nonadiabatic trajectories, state averaging with equal weights was performed over 7 states (XMS(7)-CASPT2(6,6)). All electronic-structure calculations were performed with the BAGEL program [4–5].

### 2.2. Simulation of the time-resolved photoelectron spectrum

*Theoretical approach*

The time-resolved photoelectron spectrum was computed in the classical limit of the doorway-window (DW) approximation [6–9]. A detailed description of the application of the classical DW formalism for the computation of time-resolved photoelectron spectra is given in Ref. [9] Briefly, the computational procedure includes: (i) the description of the excitation of the system by the pump pulse, (ii) the propagation of classical trajectories in the electronic ground and excited states and (iii) the description of the photoionization by the probe pulse.

(i) The pump pulse centered at $t = 0$, with carrier frequency $\omega_{pu}$ is assumed to have a Gaussian shape with envelopes $\varepsilon_{pu}(t)$ and $\tilde{\varepsilon}_{pu}(\omega)$ in the time and energy domain, respectively. A set of initial geometries **R** and momenta **P**, as well as the initial excited state are stochastically sampled from the classical doorway function [8,9]

$$D_I(\mathbf{R}, \mathbf{P}; \omega_{\mathrm{pu}}) = \tilde{\varepsilon}_{\mathrm{pu}}^2 \left(\omega_{\mathrm{pu}} - \Delta E_{IG}(\mathbf{R})\right) |\mu_{GI}|^2 \rho^{\mathrm{Wig}}(\mathbf{R}, \mathbf{P}), \qquad (1)$$

where $\Delta E_{IG}(\mathbf{R}) = E_I(\mathbf{R}) - E_G(\mathbf{R})$ is the vertical excitation energy between the ground state *G* and excited electronic state *I*, $\mu_{GI}(\mathbf{R})$ is transition dipole moment and $\rho^{\mathrm{Wig}}(\mathbf{R}, \mathbf{P})$ is the ground-state Wigner distribution.

(ii) Starting with the sampled initial conditions, surface hopping trajectories were propagated in the excited electronic states. To describe the "hole" in the ground electronic state, Born–Oppenheimer molecular dynamics simulations were performed with the same initial conditions.

(iii) The action of the probe in the excited states at the delay time $\tau$ is described by the classical window function

$$W_I(E_k, \tau; \omega_{\text{pr}}) = \sum_F \tilde{\varepsilon}_{\text{pr}}^2 \left(\omega_{\text{pr}} - E_k - IE_{I(\tau)F}(\tau)\right) \sigma_{I(T)F}(E_k, \tau), \tag{2}$$

where $\tilde{\varepsilon}_{pr}(\omega)$ denotes the Fourier transform of the probe pulse envelope $\varepsilon_{pr}(t)$ and $E_k$ the kinetic energy of the photoelectron. $IE_{I(\tau)F}(\tau) = IE_{I(\tau)F}(\mathbf{R}(\tau))$ is the ionization energy of the initial state $I = I(\tau)$ to the final state $F$ and $\sigma_{I(\tau)F}(E_k, \tau) = \sigma_{I(\tau)F}(E_k, \mathbf{R}(\tau))$ is the partial cross section. Similarly, the window function

$$W_0(E_k, \tau; \omega_{\text{pr}}) = \sum_F \tilde{\varepsilon}_{\text{pr}}^2 \left(\omega_{\text{pr}} - E_k - IE_{GF}(\tau)\right) \sigma_{GF}(E_k, \tau), \tag{3}$$

detects the "hole" in the ground state. Here $IE_{GF}(\tau)$ and $\sigma_{GF}(E_k, \tau)$ are coordinate dependent ionization energies and partial cross sections for the ionization from the ground state $G$.
The time-resolved photoelectron spectrum $P(E_k, \tau)$ is obtained by averaging over all trajectories as

$$P(E_k, \tau) = \langle D_I(\mathbf{R}, \mathbf{P}; \omega_{\text{pu}}) W_I(E_k, \tau; \omega_{\text{pr}}) \rangle - \sum_I \langle D_I(\mathbf{R}, \mathbf{P}; \omega_{\text{pu}}) W_0(E_k, \tau; \omega_{\text{pr}}) \rangle. \tag{4}$$

The first term corresponds to excited state absorption (ESA) component and the second to the ground state bleach (GSB) components of the spectrum. Finally, the time-resolved photoelectron spectrum $P(E_k, \tau)$ is expressed in terms of the binding energy, $eBE = \omega_{\text{pr}} - E_k$.

*Computational protocol and details*

We sampled stochastically a first set of 2000 coordinates and momenta from the ground-state Wigner distribution $\rho^{\text{Wig}}(\mathbf{R}_G, \mathbf{P}_G)$. By accounting for the shape of the pump pulse and the oscillator strength of the transition, $\tilde{\varepsilon}_{\text{pu}}^2 |\mu_{GI}|^2$, we obtained a set $N = 107$ initial coordinates $\mathbf{R}$ and momenta $\mathbf{P}$, as well as a set of initial excited states $I$. In our experiment the pump pulse is characterized by $\delta_{pu} = 240$ fs FWHM in the time domain and $\tilde{\delta}_{pu} = 7$ meV FWHM in the energy domain. Owing to the narrowness of the pump in the energy domain, we replaced $\tilde{\varepsilon}_{\text{pu}}^2 |\mu_{GI}|^2$ with the rectangular function $\Pi(\omega_{pu} \pm 5\tilde{\delta}_{pu})$ to increase the efficiency of the sampling.

To compute the ESA component of the photoionization spectrum the trajectories were propagated in the excited states with Tully's fewest switches surface hopping (FSSH) algorithm[10] using an in-house code [11,12]. Newton's equations for nuclear motion were propagated with the velocity-Verlet algorithm for 2000 fs. The usual time step of 0.5 fs was used. The locally diabatic formalism was used to propagate the electronic wave function and compute the hopping probabilities [13]. The energy-based decoherence procedure of Granucci and Persico [14] was employed ($\alpha = 0.1 E_h$). In the simulations 106 out of 107 trajectories were started in the first excited state ($S_1$) and one trajectory was started from the second excited state ($S_2$). At the end of the surface hopping simulations we evaluated the window function $W_I(E_k, \tau; \omega_{\text{pr}})$ and obtained the ESA signal by averaging over all trajectories. For the GSB component of the spectrum we propagated the sampled trajectories in the electronic ground state using the same velocity-Verlet algorithm and the 0.5 fs time step. We then evaluated $W_0(E_k, \tau; \omega_{\text{pr}})$ and averaged over all trajectories to obatain the GSB signal.

In our simulations the partial cross sections $\sigma_{I(\tau)F}$ and $\sigma_{GF}$ needed for the evaluation of the windows functions were approximated with Dyson orbital norms *(vide infra)* and the experimental time-resolution (Figs. 1 and S5) was accounted for by convoluting the spectrum (Eq. 4) with a Gaussian pump-probe cross-correlation function [15] of 328 fs FWHM.

*Dyson orbitals*

To simulate the time-resolved photoelectron spectrum the partial differential cross section needs to be computed

$$\frac{d\sigma_{IFk}}{d\vec{k}} = 4\pi^2 \alpha \omega |\langle \Psi_I^N | \vec{\mu} | \Phi_{Fk}^N \rangle|^2 \tag{5}$$

where $\vec{k}$ is the momentum of the photoelectron, $\alpha$ is the fine structure constant, $\omega$ is the photon energy and $\langle \Psi_I^N | \vec{\mu} | \Phi_{Fk}^N \rangle = \vec{\mu}_{IFk}$ is the dipole transition moment computed for a particular molecular geometry from the initial electronic state $I$ and the final state $F$. Expressing the wave function of the final state, $\Phi_{Fk}^N$, as an antisymmetrized product of the wave function of the ejected photoelectron, $\varphi_k$ and the bound cationic wave function $\Psi_F^{N-1}$ one obtains

$$\vec{\mu}_{IFk} = \langle \Psi_I^N | \vec{\mu} | \Phi_{Fk}^N \rangle = \langle \phi_{IF}^D | \vec{\mu} | \varphi_k \rangle \tag{6}$$

where $\phi_{IF}^D$ is the Dyson orbital describing the hole created in $\Psi_F^{N-1}$, given as

$$\phi_{IF}^D = \sum_k \langle \Psi_F^{N-1} | \hat{a}_k | \Psi_I^N \rangle \chi_k \tag{7}$$

and $\hat{a}_k$ is the annihilation operator for the molecular orbital $\chi_k$. The procedure for the computation of Dyson orbitals from CASPT2 wave function overlaps is given in Ref. [9].

In order to simulate the time-resolved photoelectron spectrum many hundreds of evaluations of the cross section are needed. Even for medium size systems as CHD accurate computations of the continuum wave functions are numerically expensive and approximations are usually done. In the sudden approximation in which the photoelectron and the cation are decoupled, the spectral strength of a transition

$$|\vec{\mu}_{IFk}|^2 = |\langle \phi_{IF}^D | \vec{\mu} | \varphi_k \rangle|^2 = \|\phi_{IF}^D\|^2 \left|\langle \overline{\phi}_{IF}^D | \mu | \varphi_k \rangle\right|^2 \approx \|\phi_{IF}^D\|^2 \tag{8}$$

is given by the square of the norm of the Dyson orbital, $\|\phi_{IF}^D\|^2$, where $\overline{\phi}_{IF}^D = \phi_{IF}^D / \|\phi_{IF}^D\|$ is the Dyson orbital normalized to one. Since in our experiment delocalized valence electrons are ejected by a high-energy probe pulse, the sudden approximation holds well.

## 2.3. Reaction path calculations

A reaction path in $C_2$ symmetry was constructed by linear interpolation in internal coordinates between the minimum of the first state of symmetry A, that is the Franck-Condon geometry (1A) and the minimum of the second state of the same symmetry, 2A. In addition to the 12 geometry points connecting the two minima, 3 geometries were extrapolated on each side of the minima resulting in a total of 19 geometries on the reaction path. The geometries were rotated to fulfil the Eckart conditions with respect to the Franck-Condon geometry.

## 2.4. Diabatization scheme

The diabatic states were obtained from the adiabatic states by a unitary transformation $\boldsymbol{\Psi}^{(d)} = \mathbf{T}\boldsymbol{\Psi}^{(a)}$. The adiabatic-to-diabatic transformation matrix $\mathbf{T}$ is computed by employing the diabatization scheme of Simah, Hartke, and Werner [16]. In this scheme one starts by defining the diabatic states to be identical to the adiabatic states at the reference Franck-Condon geometry, $\mathbf{R}_{FC}$. The transformation matrix $\mathbf{T}$ at some displaced geometry, $\mathbf{R}$, is obtained by symmetric orthogonalization of a matrix of overlaps between the electronic wave functions at the geometry $\mathbf{R}$ and the reference states $\mathbf{T} = \mathbf{S}(\mathbf{S}^\dagger \mathbf{S})^{-1/2}$, where $S_{ij} = \left\langle \Psi_i^{(a)}(\mathbf{R}_{FC}) \middle| \Psi_j^{(a)}(\mathbf{R}) \right\rangle$. To minimize the effect of geometrical changes, wave-function overlaps were calculated as a dot product of CI vectors transformed to the basis of so-called diabatic Slater determinants containing diabatic active orbitals. The diabatic orbitals at $\mathbf{R}_{FC}$ are identical to active SA-CASSCF orbitals, while at any other geometry, $\mathbf{R}$, they are obtained as symmetrically-orthogonalized projections of SA-CASSCF orbitals on the reference orbitals at $\mathbf{R}_{FC}$. CI vectors were obtained by rotation of SA-CASSCF CI vectors by the XMS-CASPT2 rotation matrix [17].

In the reaction path calculations, we included seven electronic states. From the total overlap matrix with the reference geometry (7×7) we extracted submatrices of dimension $n \times n$, which were orthogonalized and used as the adiabatic-to-diabatic transformation matrix ($\mathbf{T}$). The diabatic potentials, which are the diagonal elements of the diabatic Hamiltonian, were obtained by unitary transformation of the adiabatic Hamiltonian, $\mathbf{H}^{(d)} = \mathbf{T}\mathbf{H}^{(a)}\mathbf{T}^T$. Smooth diabatic potentials were obtained with *n = 6* for smaller $C_1$-$C_6$ distances (<1.9 Å) and with *n = 5* for larger $C_1$-$C_6$ distances (see Fig. 3).

In nonadiabatic dynamics simulations, our goal was to find the diabatic basis that best spans the active (currently populated) state of each surface-hopping trajectory at each geometry. In this way the diabatic character of the active adiabatic state, mostly the $S_1$ state, could be retrieved. For most geometries, the active state could be spanned in the basis of 4 diabatic states ($1^1A^-$, $1^1B$, $2^1A^-$ and $2^1A^+$). Therefore, along each surface-hopping trajectory we performed XMS(7)-CASPT2(6,6) single point calculations in steps of 2 fs, computed the 7×7 wave-function overlap matrix with the reference geometry, and extracted the 4×4 submatrix containing the four diabatic states of interest, the active adiabatic state and three other adiabatic states which maximize the Frobenius norm of the extracted matrix. The average population of a diabatic state, $a$, computed for a swarm of $i = 1, 2, \ldots, N$ nonadiabatic trajectories running in the adiabatic surface, $I(\tau)$, is then given as

$$P_a(\tau) = \frac{1}{N} \sum_i^N \left| T_{aI_i(\tau)}(\mathbf{R}_i(\tau)) \right|^2. \quad [9,18]$$

The contribution from the ionization of the $a$-th diabatic state in the ESA component of the photoionization spectrum was calculated as

$$P_a^{(\text{diab})} = \langle D_I(\mathbf{R}, \mathbf{P}; \omega_{\text{pu}}) W_a^{(d)}(E_k, \tau; \omega_{\text{pr}}) \rangle, \qquad (9)$$

where $W_a^{(d)}$ is the window function of the $a$-th diabatic state

$$W_a^{(d)}(E_k, \tau; \omega_{\text{pr}}) = \left| T_{aI(\tau)}(\mathbf{R}(\tau)) \right|^2 W_I(E_k, \tau; \omega_{\text{pr}}), \qquad (10)$$

and $T_{aI(\tau)}(\tau)$ is a matrix element of adiabatic-to-diabatic transformation matrix connecting the active state $I(\tau)$ and the $a$-th diabatic state.

## 2.5 References

[1] Shiozaki, T., Gyorffy, W., Celani, P. & Werner, H-J. Extended multi-state complete active space second-order perturbation theory: energy and nuclear gradients. *J. Chem. Phys.* **135,** 081106 (2011).

[2] Vlaisavljevich, B. & Shiozaki, T. Nuclear energy gradients for internally contracted complete active space second-order perturbation theory: multistate extensions. *J. Chem. Theory Comput.* **12,** 3781–3787 (2011).

[3] Park, J.W. & Shiozaki, T. Analytical derivative coupling for multistate CASPT2 theory. *J. Chem. Theory Comput.* **13,** 2561–2570 (2017).

[4] Shiozaki, T. BAGEL: Brilliantly advanced general electronicstructure library. *Wiley Interdiscip. Rev.: Comput. Mol. Sci.* **8,** el1311 (2018).

[5] Shiozaki, T. BAGEL: Brilliantly Advanced General Electronicstructure Library. http://www.nubakery.org under the GNU General Public License.

[6] Yan, Y. J., Fried, L. E. & Mukamel, S. Ultrafast Pump-Probe Spectroscopy: Femtosecond Dynamics in Liouville Space. *J. Phys. Chem.* **93,** 8149–8162 (1989).

[7] Yan, Y. J. & Mukamel, S. Femtosecond pump-probe spectroscopy of polyatomic molecules in condensed phases. *Phys. Rev. A* **14,** 6485–6504 (1990).

[8] Gelin, M. F. *et al*. Ab initio surface-hopping simulation of femtosecond transient absorption pump-probe signals of nonadiabatic excited state dynamics using the doorway-window representation. *J. Chem. Theory Comput.* **17,** 2394–2408 (2021).

[9] Piteša, T. *et al*. A combined surface-hopping, Dyson orbital and B-spline approach for the computation of time-resolved photoelectron spectroscopy signals: the internal conversion in pyrazine. *J. Chem. Theory Comput.* **17,** 5098–5109 (2021).

[10] Tully, J. C. Molecular dynamics with electronic transitions. *J. Chem. Phys.* **93,** 1061–1071 (1990).

[11] Sapunar, M., Piteša, T., Davidović, D. & Došlić, N. Highly efficient algorithms for CIS type excited state wave function overlaps. *J. Chem. Theory Comput.* **15,** 3461–3469 (2019).

[12] Piteša T., Alešković, M., Becker, K., Basarić, N. & Došlić, N. Photoelimination of nitrogen from diazoalkanes: involvement of higher excited singlet states in the carbene formation. *J. Am. Chem. Soc.* **142,** 9718–9724 (2020).

[13] Granucci, G., Persico, M., Toniolo, A. Direct semiclassical simulation of photochemical processes with semiempirical wave functions. *J. Chem. Phys.* **114,** 10608–10615 (2001).

[14] Granucci, G. & Persico, M. Critical appraisal of the fewest switches algorithm for surface hopping. *J. Chem. Phys.* **126,** 134114 (2007).

[15] Bonačić-Koutecký, V. & Mitrić, R. Theoretical Exploration of Ultrafast Dynamics in Atomic Clusters: Analysisand Control. *Chem. Rev.* **105,** 11–65 (2005).

[16] Simah, D., Hartke, B. & Werner, H.-J. Photodissociation dynamics of $H_2S$ on new coupled ab initio potential energy surfaces. *J. Chem. Phys.* **111,** 4523-4534 (1999).

[17] Finley, J., Malmqvist, P., Roos, B.O. & Serrano-Andrés, L. The multi-state CASPT2 method. *Chem. Phys. Lett.* **288,** 299−306 (1998).

[18] Landry, B. R., Falk, M. J. & Subotnik, J. E. The correct interpretation of surface hopping trajectories: How to calculate electronic properties. *J. Chem. Phys.* **139,** 211101 (2013).

# 3. Supplementary Tables

**Supplementary Table 1.** State ordering and vertical excitation energies (in eV) of the six lowest valence-excited states at the Franck-Condon geometry ($S_{0min}$) labeled according to *plus-minus* alternancy symmetry, leading configurational state functions (CSF) and their CI coefficients calculated at the XMS(7)-CASPT2[6e,6o] level of theory. For CC3 vertical excitation energies see Supplementary Table 2.

| State | $E_{exc}$ ($f_{exc}$) (XMS-CASPT2) | $E_{exc}$ (exp[a]) | Leading configurations | CI coefficients |
|---|---|---|---|---|
| $1^1A^-$ | - | - | Aufbau | 0.96 |
|  |  |  | $\pi_2\pi_2 \to \pi_1^*\pi_1^*$ | −0.14 |
| $1^1B$ | 5.18 (0.017) | 4.94 | $\pi_2 \to \pi_1^*$ | 0.97 |
| $2^1A^-$ | 5.99 (0.008) | - | $\pi_2\pi_2 \to \pi_1^*\pi_1^*$ | 0.53 |
|  |  |  | $\pi_1 \to \pi_1^*$ | −0.55 |
|  |  |  | $\pi_2 \to \pi_2^*$ | 0.46 |
| $2^1B$ | 7.50 (0.007) | - | $\sigma \to \pi_1^*$ | 0.97 |
| $1^1A^+$ | 8.31 (0.321) | 7.90 | $\pi_1 \to \pi_1^*$ | 0.69 |
|  |  |  | $\pi_2 \to \pi_2^*$ | 0.65 |
|  |  |  | $\pi_2\pi_2 \to \pi_1^*\pi_1^*$ | 0.16 |
| $2^1A^+$ | 8.66 (0.217) | - | $\pi_2\pi_2 \to \pi_1^*\pi_1^*$ | 0.68 |
|  |  |  | $\pi_2 \to \pi_2^*$ | −0.51 |
|  |  |  | $\pi_1 \to \pi_1^*$ | 0.28 |
| $3^1B$ | 10.11 (0.000) | - | $\pi_1 \to \pi_2^*$ | 0.96 |

[a] Optical and electron energy loss spectroscopies [32].

**Supplementary Table 2.** State ordering and vertical excitation energies of the lowest 6 excited states of A symmetry and lowest 4 excited states of B symmetry calculated with linear response (LR) coupled cluster singles, doubles, and triples model CC3 and the cc-pVDZ basis set at the Franck-Condon geometry of CHD. On the basis of the CI coefficients of the leading CSFs, one sees that the 2A, 6A and 7A states on LR-CC3 level correspond to the $2^1A^-$, $1^1A^+$ and $2^1A^+$ on XMS-CASPT2 level respectively.

| State | $E_{exc}$ / eV (LR-CC3) | Leading configurations | CI coefficients |
|---|---|---|---|
| 1B | 5.45 | $\pi_2 \to \pi_1^*$ | 0.94 |
| 2A | 6.54 | $\pi_2\pi_2 \to \pi_1^*\pi_1^*$ | 0.46 |
|    |      | $\pi_1 \to \pi_1^*$ | –0.41 |
|    |      | $\pi_2 \to Ry(3p_z)$ | –0.34 |
|    |      | $\pi_2 \to \pi_2^*$ | 0.22 |
| 3A | 7.26 | $\pi_2 \to Ry(3s)$ | 0.90 |
| 2B | 7.29 | $\sigma_2 \to \pi_1^*$ | 0.91 |
| 4A | 7.71 | $\sigma_1 \to \pi_1^*$ | 0.92 |
| 5A | 8.19 | $\pi_2 \to Ry(3p_z)$ | 0.84 |
|    |      | $\pi_1 \to \pi_1^*$ | 0.29 |
|    |      | $\pi_2 \to Ry(3s)$ | 0.20 |
|    |      | $\pi_2\pi_2 \to \pi_1^*\pi_1^*$ | 0.18 |
| 3B | 8.34 | $\pi_2 \to Ry(3p_x)$ | 0.90 |
|    |      | $\pi_1 \to Ry(3s)$ | 0.23 |
| 6A | 8.82 | $\pi_1 \to \pi_1^*$ | 0.53 |
|    |      | $\pi_2 \to \pi_2^*$ | 0.67 |
|    |      | $\pi_2\pi_2 \to \pi_1^*\pi_1^*$ | 0.29 |
| 4B | 8.94 | $\pi_2 \to Ry(3p_y)$ | 0.90 |
| 7A | 9.09 | $\pi_2\pi_2 \to \pi_1^*\pi_1^*$ | 0.44 |
|    |      | $\pi_2 \to \pi_2^*$ | –0.56 |
|    |      | $\pi_1 \to \pi_1^*$ | 0.35 |

# 4. Supplementary Figures

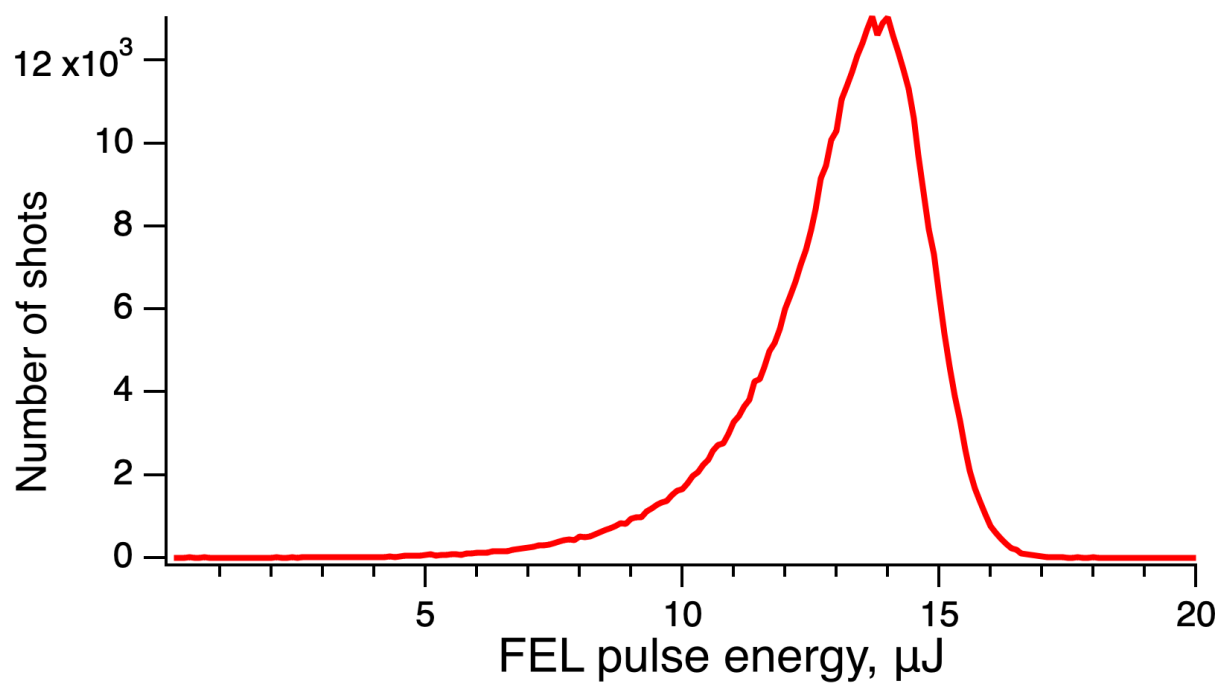

**Supplementary Figure 1.** Distribution of the FEL pulse energy for about 400000 shots during one pump-probe delay scan measurement. The data are sampled with increments of 0.1 µJ.

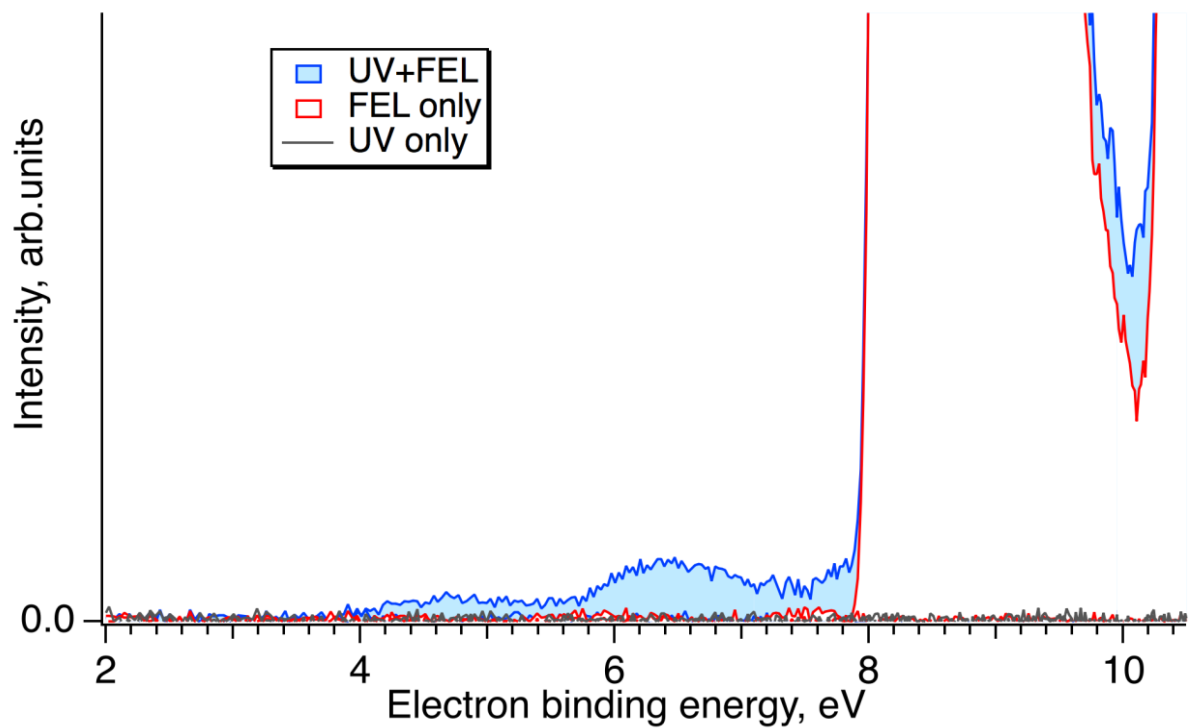

**Supplementary Figure 2.** Photoelectron spectra recorded for (1) the time delay $t_0$, corresponding to the overlap of the UV-pump and FEL-probe pulses – blue curve; (2) UV pulse only – grey curve and (3) FEL pulse only – red curve.

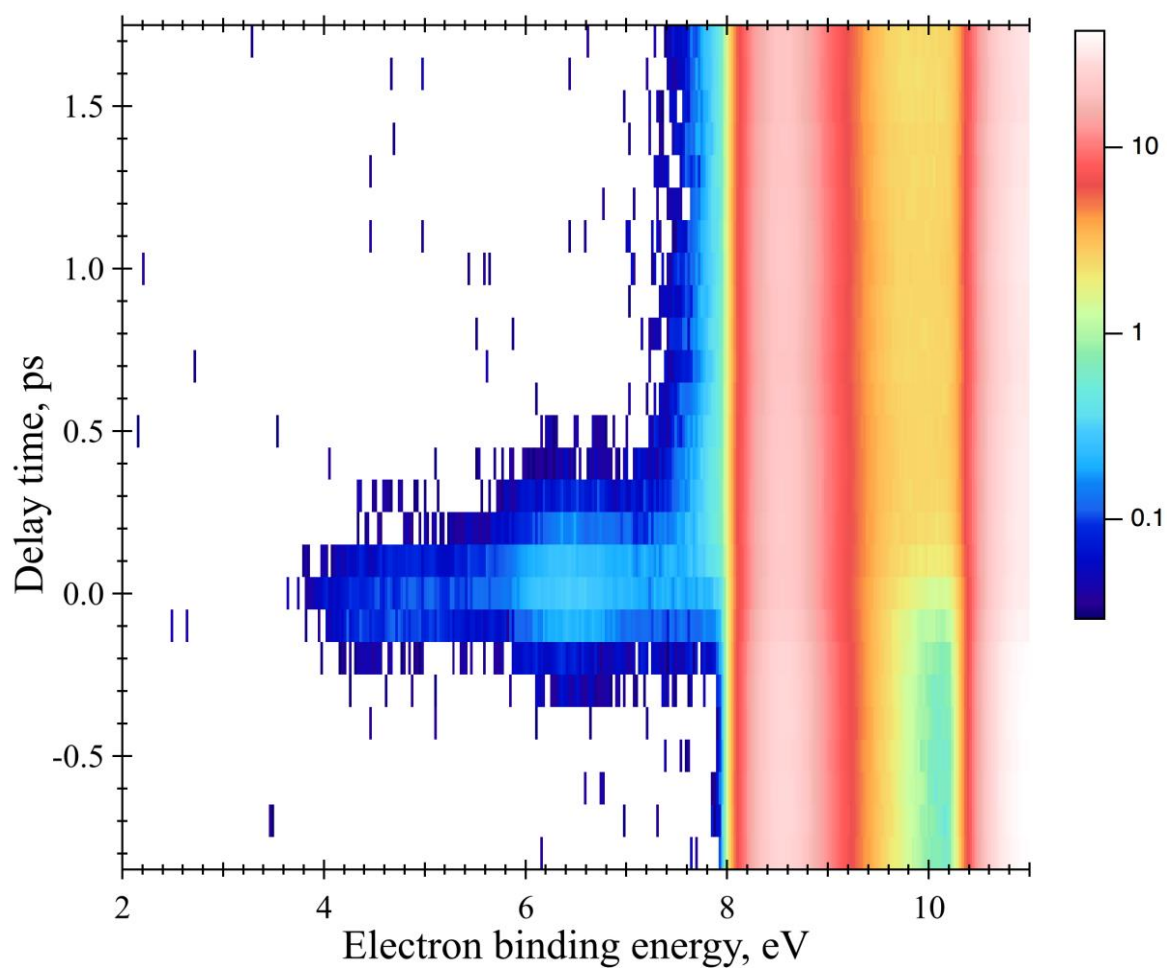

**Supplementary Figure 3.** Evolution of the photoelectron spectra as a function of the pump-probe delay, measured with a step of 100 fs. The 2D map shown here is without subtraction of the ground-state spectrum and was used to construct Figure 1 of the main paper.

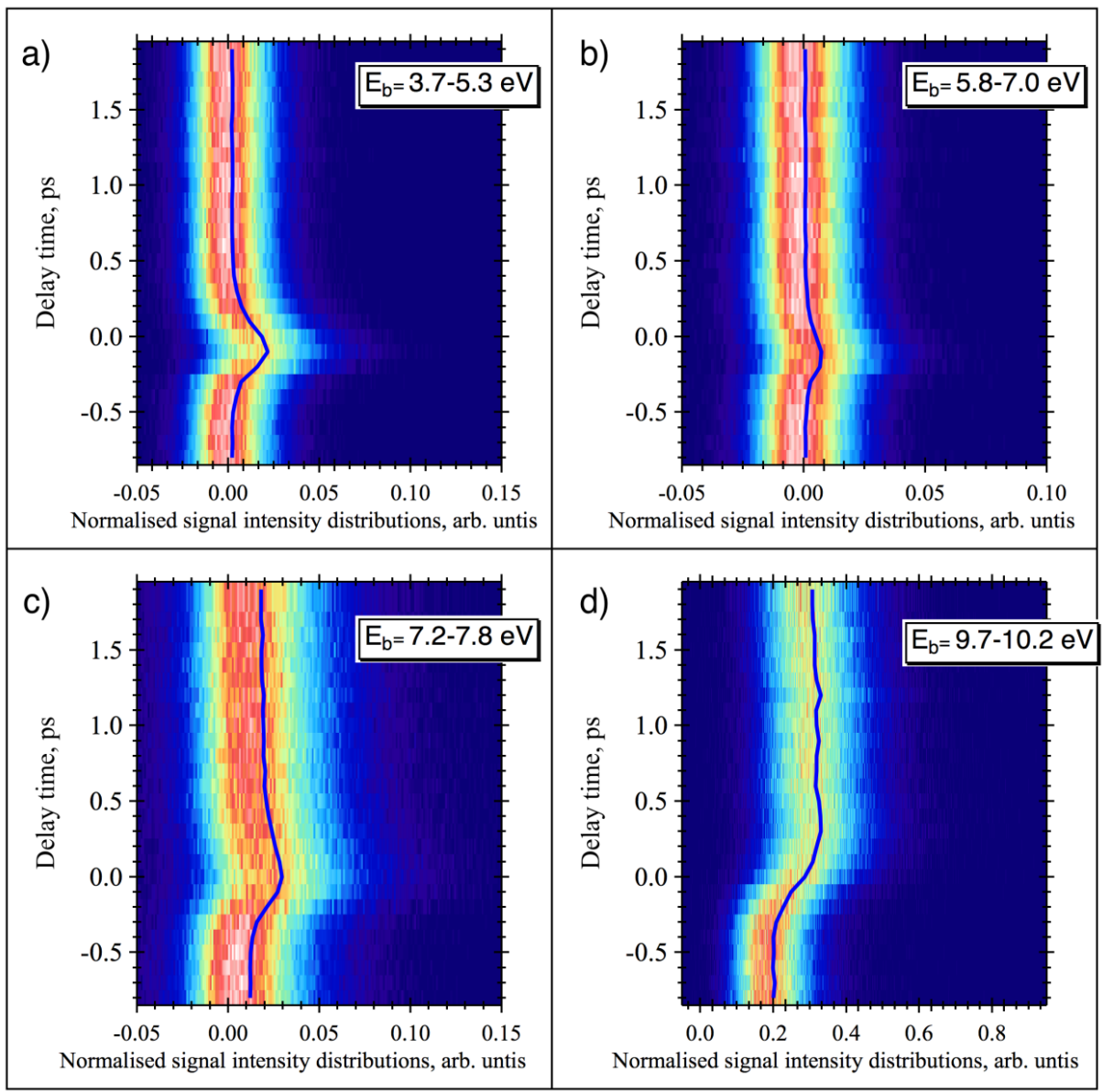

**Supplementary Figure 4.** Normalised time-of-flight signal intensity distributions, integrated over the photoelectron TOF regions, which correspond to the photoelectron binding energy regions of about a) 3.7-5.3 eV, b) 5.8-7.0 eV, c) 7.2-7.8 eV and d) 9.7-10.2 eV. The histograms of the intensity distributions were obtained for each delay point of the pump-probe delay scan of one dataset. The blue curves correspond to the median intensity distributions and show the evolution of the photoelectron signal as a function of the pump-probe delay time.

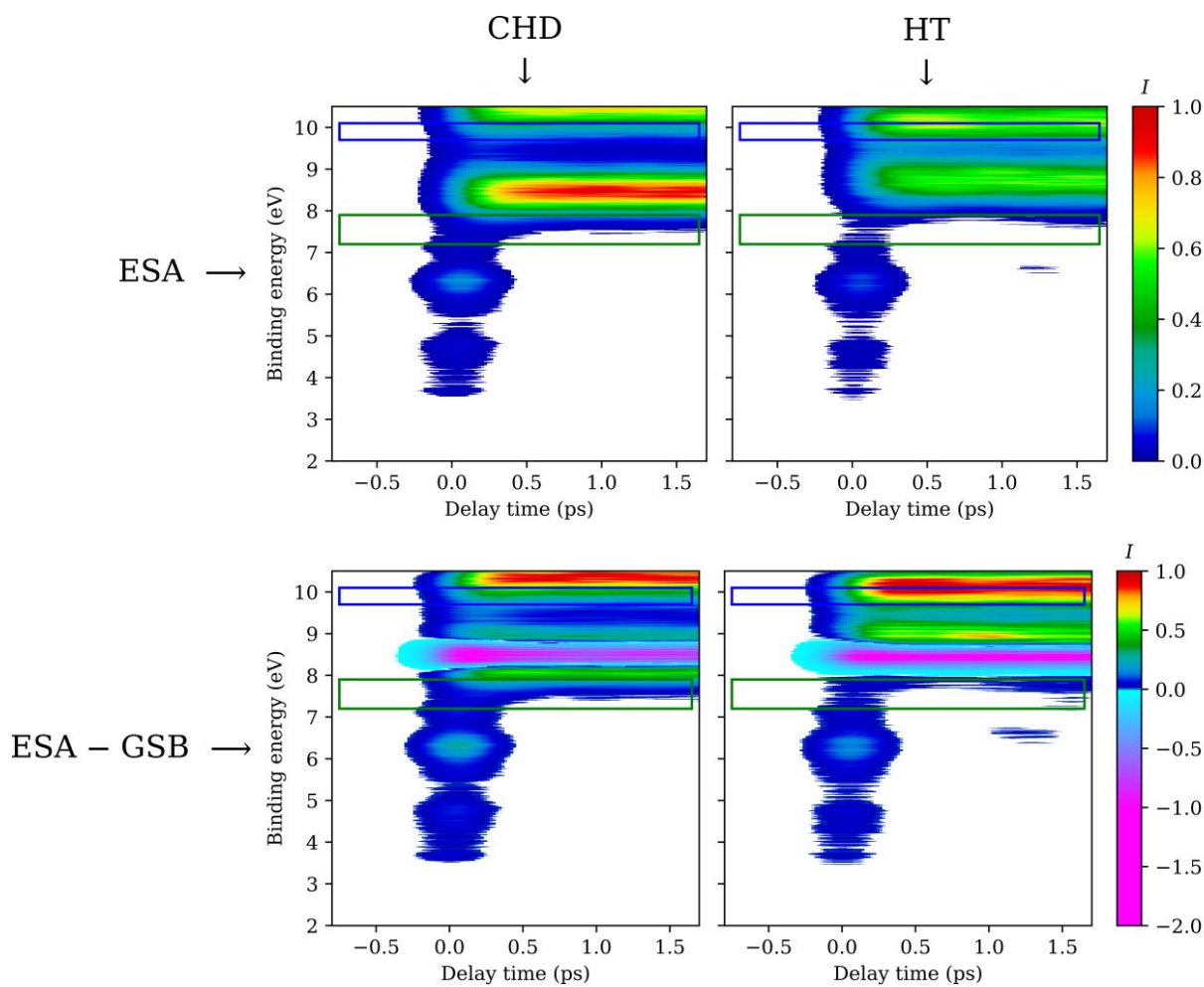

**Supplementary Figure 5.** Upper panels: Simulated time-resolved photoelectron spectra of pump-excited geometries propagated in the excited states using classical trajectory surface hopping dynamics and ending as CHD (left) or HT (right). The spectrum is computed as the excited-state absorption (ESA) component of the transient absorption pump-probe signal for excited states in the continuum (for details, see SI sec. 2.2) Lower panels: Time-resolved photoelectron spectra obtained by subtracting the ground-state bleach (GSB) contribution from the ESA contribution. The GSB contribution is computed by propagating the pump-excited geometries in the electronic ground state. Bands C and D are marked with green and blue rectangles, respectively.

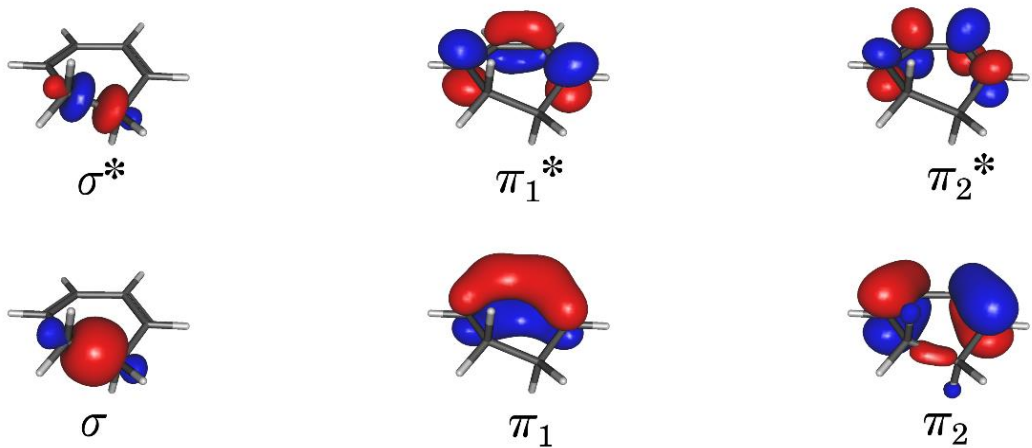

**Supplementary Figure 6.** The orbitals constituting the (6e,6o) active space at the ground state minimum energy geometry of CHD (C$_2$ symmetry).

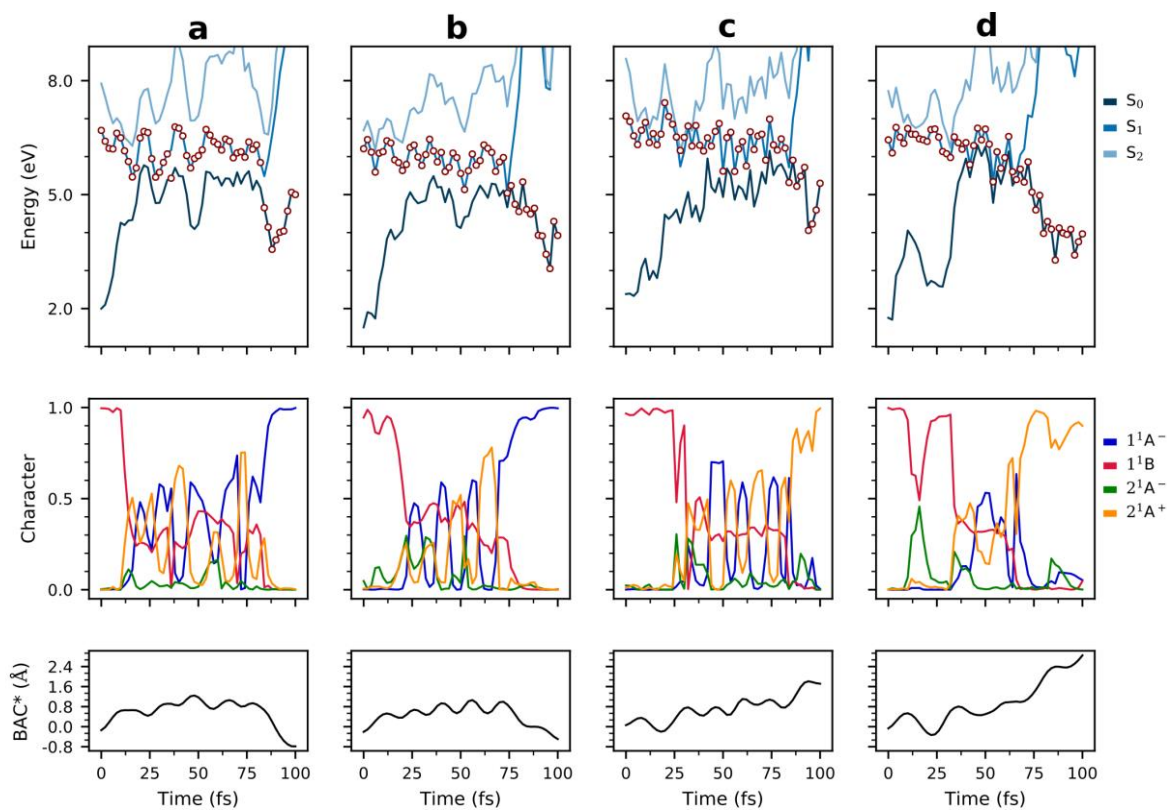

**Supplementary Figure 7. a,b** Nonadiabatic dynamics trajectories leading to CHD. **c,d** Nonadiabatic dynamics trajectories leading to HT. **Top**, Time evolution of the potential energy of the electronic ground state $S_0$ (dark blue) and the two lowest excited states $S_1$ (blue) and $S_2$ (light blue). Dots mark the currently populated electronic state. **Middle**, Decomposition of the $S_1$ state in terms of four diabatic states, $1^1A^-$ (blue), $1^1B$ (red), $2^1A^-$ (green) and $2^1A^+$ (orange) along the trajectories. **Bottom**, Evolution of the extended bond alternation coordinate (BAC*) along the nonadiabatic trajectories.

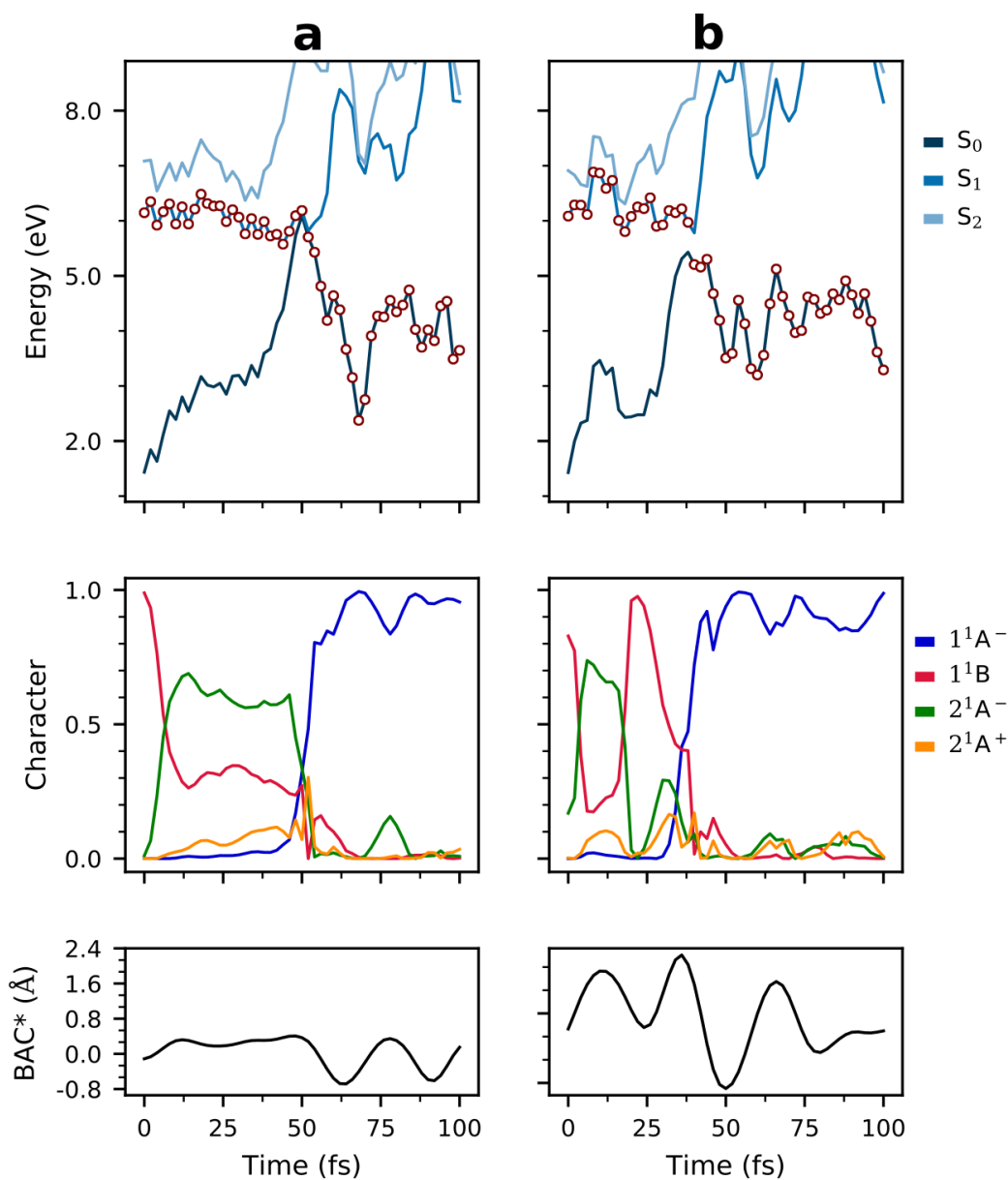

**Supplementary Figure 8. a,b,** Nonadiabatic dynamics trajectories with short R($C_1$-$C_6$) bond distance and small BAC* at the time of deactivation to the ground state. **Top**, Time evolution of the potential energy of the electronic ground state $S_0$ (dark blue) and the two lowest excited states $S_1$ (blue) and $S_2$ (light blue). Dots mark the currently populated electronic state. **Middle**, Decomposition of the $S_1$ state in terms of four diabatic states, $1^1A^-$ (blue), $1^1B$ (red), $2^1A^-$ (green) and $2^1A^+$ (orange) along the trajectories. Notice the small contribution of the $2^1A^+$ state during the dynamics. The hop to the ground state occurs from the $2^1A^-$ state (a) and $1^1B$ (b) state.

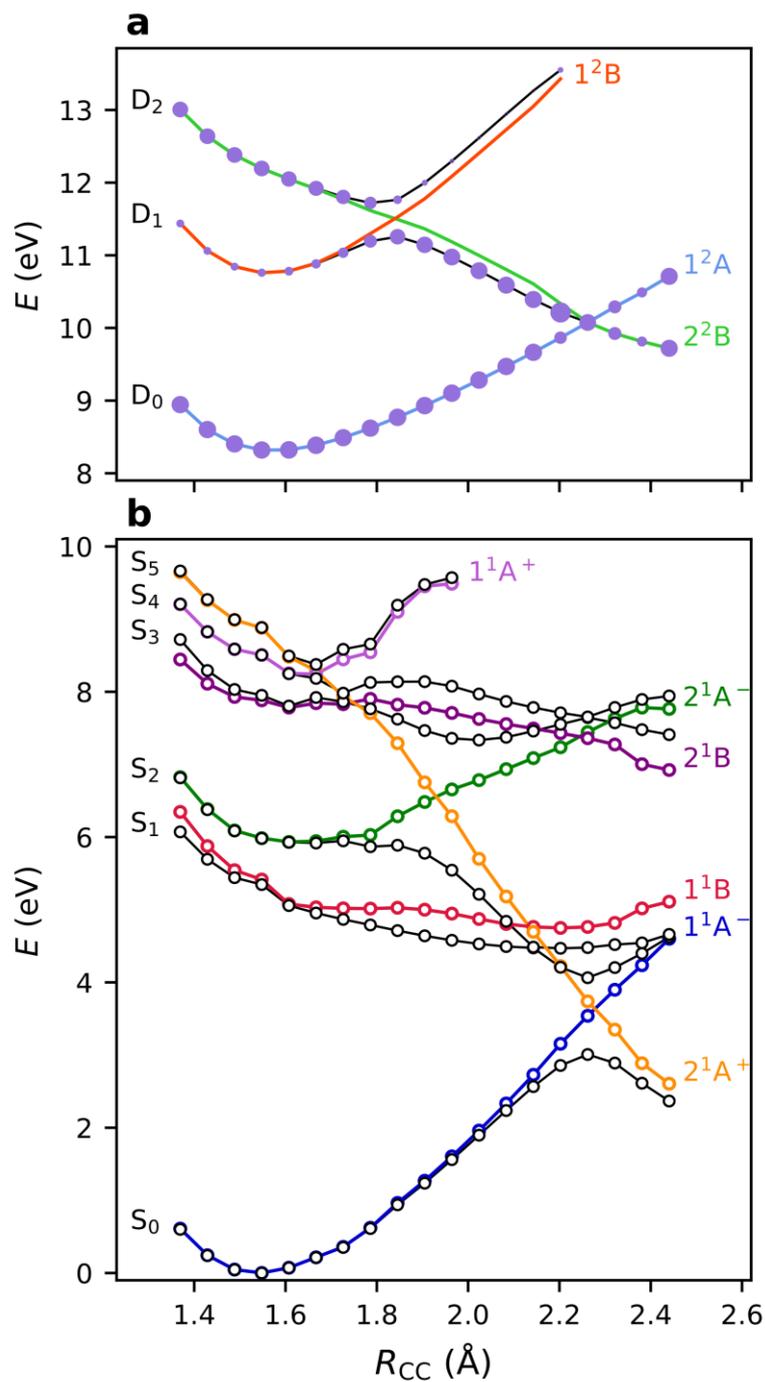

**Supplementary Figure 9.** Adiabatic and diabatic potential energy surfaces of (**a**) $CHD^+$ and (**b**) CHD along the linearly interpolated ring-opening reaction coordinate in $C_2$ symmetry. The size of the circles on the cationic adiabatic surfaces is proportional to the square of the Dyson norm from the $S_1$ adiabatic state of CHD to the corresponding doublet state of the $CHD^+$ cation.